\documentclass[twocolumn,tighten,times]{aastex701}
\usepackage{amsmath}
\usepackage{svg}
\usepackage{graphicx}
\usepackage{comment}
\usepackage[ruled]{algorithm2e}
\usepackage{tabularx}
\usepackage{colortbl}
\usepackage{diagbox}
\usepackage{slashbox}
\usepackage{multirow}
\usepackage{cleveref}

\newcommand{\x}{{x_{ij}}}

\newcommand{\disappear}[1]{}
\newcommand{\modelname}{\textsc{Neo}}

\begin{document}
\title{\modelname{}: Photometric Super-Resolution for Improving Galaxy Morphological Measurements using Conditional Generative Adversarial Networks}
\author[0009-0007-7313-3735]{Samuel Kahn}
\affiliation{Department of Astronomy and Astrophysics, University of California,
            Santa Cruz, 1156 High Street, Santa Cruz, CA 95064 USA}
\email{shkahn@ucsc.edu}

\author[0000-0002-8543-761X]{Ryan Hausen}
\affiliation{Data Science and AI Institute, The Johns Hopkins University,
            3400 N. Charles St., Baltimore, MD 21218, USA}
\email{rhausen@jhu.edu}

\author[0000-0001-9935-9109]{Hubert Bretonnière}
\altaffiliation{Current address: Vall d´Hebron Institute of Oncology (VHIO), Centro Saturnino, c/Saturnino Calleja 11-13, 08035, Barcelona, Spain}
\affiliation{Department of Astronomy and Astrophysics, University of California,
            Santa Cruz, 1156 High Street, Santa Cruz, CA 95064 USA}
\email{hubertbretonniere@gmail.com}

\author[0000-0003-4761-2197]{Nicole E. Drakos}
\affiliation{Department of Physics and Astronomy, University of Hawaii, Hilo, 200 W Kawili St, Hilo, HI 96720, USA}
\email{ndrakos@hawaii.edu}

\author[0000-0002-4271-0364]{Brant E. Robertson}
\affiliation{Department of Astronomy and Astrophysics, University of California,
            Santa Cruz, 1156 High Street, Santa Cruz, CA 95064 USA}
\email{brant@ucsc.edu}

\begin{abstract}
The measurement of galaxy morphological parameters from astronomical images
features in a wide range of modern analyses, including galaxy evolution
and  cosmological weak lensing studies.
The precision and accuracy of morphological parameter estimation can be influenced
by several key factors. The effective seeing of the image, summarized by the point
spread function (PSF), limits how galaxy features or light profiles are resolved.
The pixel scale of the detector also influences the resolution and the amount
of statistical information available for a given object. The depth of the observations
determines the signal-to-noise ratio of the image.
Improving each of these factors is very costly, either in terms of detector upgrades,
observatory design, or observing time. Here, we develop
a conditional generative adversarial network, called \modelname{}, trained
to transform existing ground-based images into sharper, finer-scale images
comparable to space-based image quality.
We demonstrate that \modelname{} improves the accuracy of measured morphological
parameters by factors of $2$--$10$ when trained
to translate Subaru Hyper Suprime-Camera (HSC) images
to approximate Hubble Space Telescope (HST) data.
\modelname{} is designed for applicability to ongoing,
large-scale surveys such as the Legacy Survey of Space and Time (LSST)
conducted by Vera C. Rubin Observatory in combination with
space telescopes such as HST, James Webb Space Telescope, and Nancy
Grace Roman Space Telescope.
These results suggest that \modelname{} could be used to improve
both cosmological and galaxy evolution analyses based on massive, ground-based
survey datasets like LSST.
The model code is open source and available at \url{https://purl.archive.org/neo/code}.
\end{abstract}
\keywords{\uat{Galaxies}{573} --- \uat{Galaxy Classification Systems}{582} --- \uat{Hubble Space Telescope}{761}}

\section{Introduction}
\label{sec:introduction}

Galaxies exhibit a diverse range of morphologies, such as the familiar spiral,
elliptical, or irregular classifications
that reflect the physics that drive their evolution over cosmic time
\citep[e.g.,][]{conselice2014a,MHC_2015,willett2017a,hausen2020}.
Morphological parameters of galaxies, such as
ellipticity, size, and \citet{sersic1963a} index, quantify
trends in the surface brightness distributions of galaxies with different
morphological classes \citep[e.g.,][]{yang2025a,carreira2026a}.
Morphological parameters also reflect the
physical evolution of galaxies, as they vary between star-forming and quiescent
galaxies \citep{franx2008,bell2012,mortlock2013} and change with
redshift \citep{vanderwel2011,vanderwel2014a,guo2015,huertas-company2024a,miller2025a,allen2025a}. Galaxy morphology correlates
strongly with galaxy formation history \citep[e.g.,][]{brennan2017},
which can provide an avenue for testing galaxy formation theories
empirically. Observed galaxy shapes, sizes,
and position angles also depend on large-scale cosmic structures, as
matter intervening between background source galaxies and the observer
bends the path of light.
Analyzing this weak gravitational lensing signal that leads to
coherent distortions in the observed shapes of background
galaxies helps to constrain both cosmological parameters and
the galaxy--halo connection  \cite[e.g.,][]{hoekstra2008,troxel2018a,hikage2019a,heymans2021a,des2022a,dalal2023a,des2026a}.

High-resolution imaging typically serves for morphological classification of
galaxies, with many catalogs originating from HST data
\citep[e.g.,][]{abraham1996,scarlata2007,kartaltepe2012,kartaltepe2015a,Laigle_2016}. Surveys
from the James Webb Space Telescope \citep[JWST; e.g.,][]{kartaltepe2022,casey2022,bagley2024a,finkelstein2025a,eisenstein2026a}, Euclid \citep{euclid_reference_paper}, the
Vera C. Rubin Observatory Telescope \citep[hereafter Rubin]{ivezic2019}, and the
Roman Space Telescope \citep{wfirst, akeson2019} will significantly enhance data
quality for morphological classifications. This increase in high-resolution data
has driven the development of machine-learning algorithms for efficient galaxy
morphology classification \citep[e.g.,][]{snyder2019, hausen2020, vega_2021,
rose2022,robertson2023a}.

When high-resolution imaging is unavailable,
super-resolution techniques offer viable alternatives for
obtaining morphological measurements \citep{shibuya2022},
and a range of techniques have been employed to augment
astronomical observations.
Generative Adversarial Networks (GANs), where AI models
learn to create artificial images that approximate real
scenes, can recover galaxy features from artificially degraded images \citep{Schawinski_2017}.
Other examples include the successful application of supervised deep learning to galaxy–galaxy lens finding \citep[e.g.,][]{Lanusse_2017} and the
training of convolutional neural network (CNN) variants for super-resolution applied to wide-field radio astronomy to reveal source morphologies \citep{Dabbech_2022}.
Super-resolution and denoising techniques applied to X-ray images from the European Space Agency's XMM-Newton telescope have increased the scientific value of the XMM-Newton archive by uncovering previously undetected features \citep{Sweere_2022}. \cite{Sukurdeep2023Super} introduced a framework for multi-frame image deconvolution and super-resolution, demonstrating high-fidelity reconstructions of artificially degraded Subaru Hyper Suprime-Camera (HSC) images. In  \cite{park2024deepersharperfasterapplication}, HST images were super-resolved to JWST resolution using a Transformer network \citep{vaswani17} pre-trained on synthetic galaxy images, showing significant enhancements of morphological measurements toward JWST-level resolution. \cite{hettiarachchi2024} introduced a GAN to upscale low-resolution galaxy images, improving the peak signal-to-noise ratio and structural similarity index metric of the galaxies. To improve measurements in solar physics, \cite{rahman2025} developed a conditional GAN \citep{cgan} to denoise solar observations, recovering spatial and spectral information comparable to the original ground-truth images.

An important potential application of AI/ML in astronomy
concerns improving the resolution of astronomical images.
Initial
work was performed by
\citet{gan2021seeinggan} who developed {\it SeeingGAN} to transform images of single HSC sources into HST-like images.
Their method successfully reproduced structural similarity metrics adapted from the computer vision field,
and provided an important first step toward super-resolution modeling of ground-based images.
Building on these efforts, here we introduce
\modelname{}, a neural network designed to super-resolve
ground-based data to the resolution of space-based images over wide areas.
By applying \modelname{} to HSC data, we
demonstrate that the model achieves
remarkable accuracy in morphological parameters compared
to high-resolution HST data in the same field.
Our work makes further advances on prior models by
enabling super-resolution analysis on entire mosaic fields
rather than on small regions or single sources, and by
evaluating the efficacy of the model
through quantitative metrics that assess
how astronomical measurements on the super-resolution
images improve over native-resolution, ground-based imagery.

The presentation of \modelname{} is organized
as follows:
Section~\ref{sec:model} details the model structure and methodology.
Section~\ref{sec:dataset} describes the training datasets
from HSC and HST.
Section~\ref{sec:results} details quantitative results
from the application of \modelname{} to the astronomical datasets.
Section \ref{sec:discussion} discusses the results in the
context of prior work. We conclude and summarize our work in
Section~\ref{sec:conclusion}.
We provide two appendices that present more
technical details
for the interested reader. Appendix \ref{Appendix A}
identifies some failure modes that illustrate the
limitations of our modeling. Appendix \ref{Appendix B} analyzes the noise properties of the super-resolution images and evaluates object fluxes measured from them.
Where appropriate, we adopt the \citet{oke1983a} AB magnitude system.

\section{\modelname{}: A Model for Astronomical Super-Resolution}
\label{sec:model}

\modelname{} is a novel neural network architecture that performs image-to-image translation and super-resolution simultaneously.
\modelname{} is designed to translate images
from ground-based quality seeing to space-based quality
and resolution.
For the current work, \modelname{} has been trained to
translate between HSC and HST images, but the model can in principle be applied to any low-resolution/high-resolution pair
of data and retrained. For instance, the \modelname{} model and architecture can be applied to enhance surveys like LSST when paired with data from space telescopes like Euclid or Roman.
The rest of this Section describes the \modelname{} architecture.

\subsection{Architecture Overview}
\label{sec:model:overview}

The \modelname{} architecture is inspired by the U-Net \citep{unet}
and
Super-Resolution Generative Adversarial Network \citep[SRGAN]{srgan}
architectures.
Since \modelname{} attempts to translate between low-resolution and
high-resolution datasets by generating a super-resolution model
image, the architecture of the model roughly follows the
common U-Net structure.
The model begins by pre-processing cutouts from the data to rescale the pixel
values and ensure each local image sample is padded to a multiple
of two in pixel dimensions. Padding the input image to a multiple of two ensures
that the contraction (Section \ref{sec:model:contraction}) and expansion
(Section \ref{sec:model:expansion}) phases of the model can be applied
without issues related to non-integer pixel dimensions.
The model continues with a contraction phase, where the
input image properties are translated to a lower-dimensional
latent image, followed by an
expansion phase where the image is re-grown in spatial extent
to the original image size.
Next, the expanded image is supplied to a super-resolution phase
that models the image features below the original image resolution
scale.
The model images are then post-processed to invert the scaling
and padding applied in the pre-processing step before being
integrated into an output model mosaic at the higher spatial
resolution.
A graphical representation of the architecture is
provided in Figure \ref{fig:generator}, and
the rest of this section describes each
phase in detail.

\subsubsection{Pre-Processing Phase}
\label{sec:model:preprocess}

Normalization of the range of data values
has been shown to improve the training stability and performance
of neural networks \citep{huang2023}.
For photographic or network graphic images, such as PNG
or JPEG images, the pixel values
are normalized to a range of [0,1] or [0,255].
Normalization applied to such images simply rescales and
shifts the
bounded distribution of pixel values.
Astronomical images, however, are not
normalized over a range of bounded values.
Further, to the extent possible, the output of the model should
recover the expected units and scaling of the target data distribution.
To that end, we pre-process cutouts from the input data images to prepare them
for use with the \modelname{} architecture. We present the pre-processing
steps below. After the model is applied,
the resulting model images are then post-processed to recover the original
range of data values and cropped to the original spatial extent of the input data sample (see Section \ref{sec:model:postprocess} below).

Each input mosaic
image cutout has a size of $100^2$ pixels and is pre-processed in several steps, as represented by the {\it Scale-Pad} layer in
Figure \ref{fig:generator}.
First, the pixel values in
each image are clipped to a minimum of 0 and a maximum of the
99.999$^{\textrm{th}}$ percentile value in the full mosaic images. The range of the clipped
image is then compressed using a log scaling inspired by a method used by the DS9 tool
\citep{ds9},

\begin{equation}
\label{eq:scale}
\textrm{Scale}_{\alpha, b}(x) = \frac{\log(\alpha x + 1)}{\log(\alpha + 1)} - b.
\end{equation}

\noindent
In the model presented here, we adopt $\alpha=10^3$ and $b=1$.
After clipping and scaling, the input
image pixel values span a range of $[-1, 1]$.
Finally, the model applies a reflective padding of
14 pixels along each side of the image.
The input image cutout in \modelname{} is $128\times128$ pixels, with
each dimension divisible by two.

\subsubsection{Contraction Phase}
\label{sec:model:contraction}

Figure \ref{fig:generator} shows the contraction phase of the model as
a series of layers colored in blue.
After pre-processing, the data flows into a series of
seven \textit{contraction operators}
$\psi_{f,\beta} : \mathbb{R}^{w \times h \times c} \mapsto \mathbb{R}^{w/2 \times h/2 \times c}$, where $w$ and $h$ are the width and height of
the image data provided to the operator and $c$ represents the
number of image channels at that location in the model.
Mathematically, we can represent $\psi_{f,\beta}$ as a function
of the input data $\mathbf{X}$ as

\begin{equation}
\label{eq:contraction-op}
\psi_{f,\beta}(\textbf{X}) =
\begin{cases}
    \mathcal{A}_{0.2}(C_{f,4,2,1}(\textbf{X})),     & \text{if $\beta=0$}\\
    \mathcal{A}_{0.2}(B(C_{f,4,2,1}(\textbf{X}))),  & \text{if $\beta=1$}
 \end{cases}
\end{equation}

\noindent
where
\begin{equation}
\label{eq:leakyrelu}
\mathcal{A}_\alpha=\max(x, \alpha x)
\end{equation}

\noindent
is the leaky rectified linear unit (leakyReLU) function, $B$ is the batch normalization function
\citep{ioffe2015a}, and  $C_{f,k,s,p}$, is a convolutional function with $f$
filters, kernel size $(k, k)$, stride of $(s, s)$,
and the input zero padded by $(p, p)$. The combination of the kernel size,
stride, and padding paramters chosen for the convolution function results in a
reduction of the width and height of the input by a factor of two.
At the end of the contraction phase, the latent images have
spatial dimensions of $w$ = 1 and $h$ = 1.

\subsubsection{Expansion Phase}
\label{sec:model:expansion}

Once the data leaves the contraction phase, the image spatial
extents are regrown from the latent images by a series of
seven \textit{expansion operators}
$\mathcal{E}_f: \mathbb{R}^{w \times h \times c} \mapsto \mathbb{R}^{2w \times 2h \times f}$. We can describe these expansion
operators mathematically as a function of the input data $\mathbf{X}$ as

\begin{equation}
\label{eq:expansion-op}
    \mathcal{E}_f(\textbf{X}) = \mathcal{A}_0(B(T_{f,4,2,1}(\textbf{X})))
\end{equation}

\noindent
where $\mathcal{A}_0$ is the ReLU function (i.e., $\alpha=0$),
$B$ is the batch normalization function, and
$$T_{f,k,s,p}(\textbf{X}) = C_{f,k,1,k-1-p}(U_{s}(\textbf{X}))$$
is the transpose convolution function \citep{zeiler2010} parameterized by $f$
filters, kernel size $(k, k)$, stride of $(s, s)$, and padding
$(p, p)$, where
$$U_s: \mathbb{R}^{w \times h \times c} \mapsto \mathbb{R}^{[(w-1) \cdot s + 1] \times [(h-1) \cdot s + 1] \times c}$$
is an upscaling function that inserts $(s-1)$ zeros between entries
in the input along the width and height dimensions and $C$ is the
convolution function. Our implementation uses $s=2$, which
results in a transformation $U_2: \mathbb{R}^{w \times h \times c} \mapsto \mathbb{R}^{[2w-1] \times [2h-1] \times c}$. Setting $k=4$ and $p=1$ for the
transpose convolution function results in the following
parameterization of the convolution function:
$C_{f,4,1,2}: \mathbb{R}^{w \times h \times c} \mapsto \mathbb{R}^{[w+1] \times [h+1] \times f}$.
Both operations together result in the doubling of the height
and width of the input tensor.
The outputs from the
contraction phase are concatenated with the outputs in the expansion phase along
the channel dimension when the width and height dimensions
of the images match. The first three
expansion operator layers
are regularized using the dropout technique \citep{dropout}, with
50\% of the connections randomly censored.
Once the seven expansion operators have been applied, the spatial
extent of the images matches those of the original input after the Scale-Pad operator has been applied.

\subsubsection{Super-Resolution Phase}
\label{sec:model:superresolution}

Once the expansion phase of the model has regrown an image
to the original spatial extent,
the super-resolution phase of the model generates the
sub-pixel information below the pixel scale of the
input data.
The super-resolution phase is composed of two
\textit{super-resolution operators}
$\mathcal{S}_{\beta} : \mathbb{R}^{w \times h \times c} \mapsto
\mathbb{R}^{\beta w \times \beta h \times c}$.
As a function of the input data $\mathbf{X}$, the
super-resolution operator can be written as
\begin{equation}
\label{eq:superres-op}
    \mathcal{S}_{\beta}(\textbf{X})=\mathcal{R}(P_\beta(\textbf{X}))
\end{equation}
\noindent
where $\mathcal{R}$ is the parametric rectified linear unit function
\citep{prelu}, $\mathcal{R}(x)=\max(0,x) + \alpha * \min(0, x)$ with $\alpha$
as a learnable parameter and $P_\beta$ is the subpixel convolution
function \citep{subpixel} with a scaling factor of $\beta$.
 The subpixel convolution projects feature maps from the channel dimension into
spatial dimensions, increasing image resolution without explicit
interpolation.
The two super-resolution operators apply
subpixel convolution with $\beta=3$ and $\beta=2$ sequentially.
The final operational layer of the super-resolution phase applies a
convolution function $C_{1,1,1,0}$ to the data,
followed by a hyperbolic tangent function
applied element-wise.
The resulting super-resolution model image
then has a spatial extent increased by $6\times6$ relative to the
original $h\times w$ of the input data, after cropping.

\subsection{Data Post-Processing}
\label{sec:model:postprocess}

\begin{figure*}
\centering
\includegraphics[width=\textwidth]{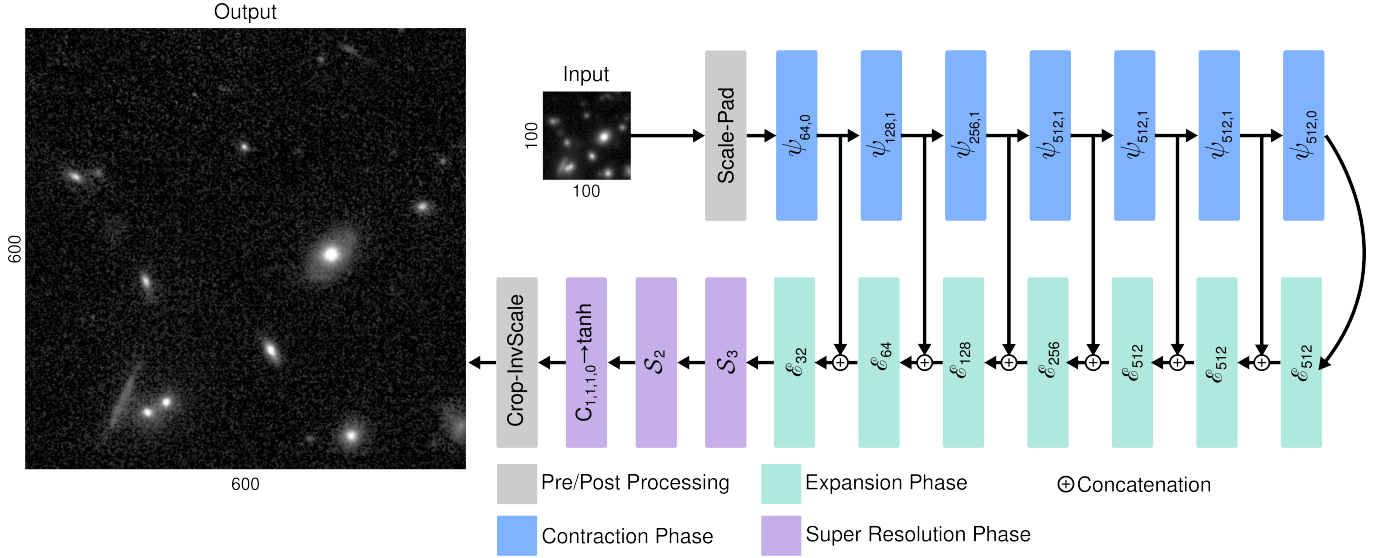}
\caption{Schematic of the \modelname{} generator network including
         pre-processing and post-processing steps.
         In the \modelname{} neural
         network, the generator is defined by a \textit{contraction phase},
         \textit{expansion phase}, and \textit{super-resolution
         phase}. The contraction
         operators $\psi_{f,b}$ reduce the spatial extent of the tensors by a
         factor of two, and are described in Section \ref{sec:model:contraction}. The expansion
         operators $\mathcal{E}_f$ grow the spatial extent of the tensors
         by a factor of two, and are described in Section \ref{sec:model:expansion}. The
         super-resolution operators $\mathcal{S}_\beta$ perform the effective subpixel
         modeling, and are described in Section \ref{sec:model:superresolution}.
         The output of the \modelname{} neural
         network is a $600^2$-pixel super-resolution image.
         ``Scale-Pad'' in the
         diagram refers to the pixel value scaling and reflective padding described
         in Section \ref{sec:model:preprocess}. ``Crop-InvScale'' in
         the diagram refers to the post-processing operations where reflective
         padding that was added during ``Scale-Pad'' is cropped away and the
         scaling is inverted as described in Section
         \ref{sec:model:postprocess}.}
\label{fig:generator}
\end{figure*}

At the end of the super-resolution phase, the model image cutout
generated by \modelname{} has dimensions of $768\times768$ pixels,
with pixel values
spanning a range of $[-1, 1]$.
The \modelname{} then post-processes these model image cutouts
to invert the scaling applied in pre-processing and to crop the cutouts
to the original on-sky extent.
The following steps are applied as post-processing
steps, which are represented by the {\it Crop-InvScale} layer in Figure \ref{fig:generator}.
First, the center $600\times600$ region of the model image
cutout is cropped to remove the up-sampled
reflective padding. The pixel values are rescaled
using a transformation to invert the original scaling used in data
preparation during pre-processing. Mathematically, this
transformation is written as

\begin{equation}
\label{eq:invscale}
\textrm{InvScale}_{\alpha, b}(x) = \frac{(\alpha + 1)^{x + b} - 1}{\alpha},
\end{equation}

\noindent
where $\alpha=10^3$ and $b=1$ to reflect the scaling applied in
pre-processing. At this point, the image does not contain
any negative values from the background subtracted image that were clipped
during pre-processing.
Negative noise values are added back in by
randomly sampling from the positive noise values and
subtracting those noise samples from the zero-valued pixels.

\subsubsection{Discriminator Network}
\label{sec:model:discriminator}

\begin{figure*}
\centering
\includegraphics[width=\textwidth]{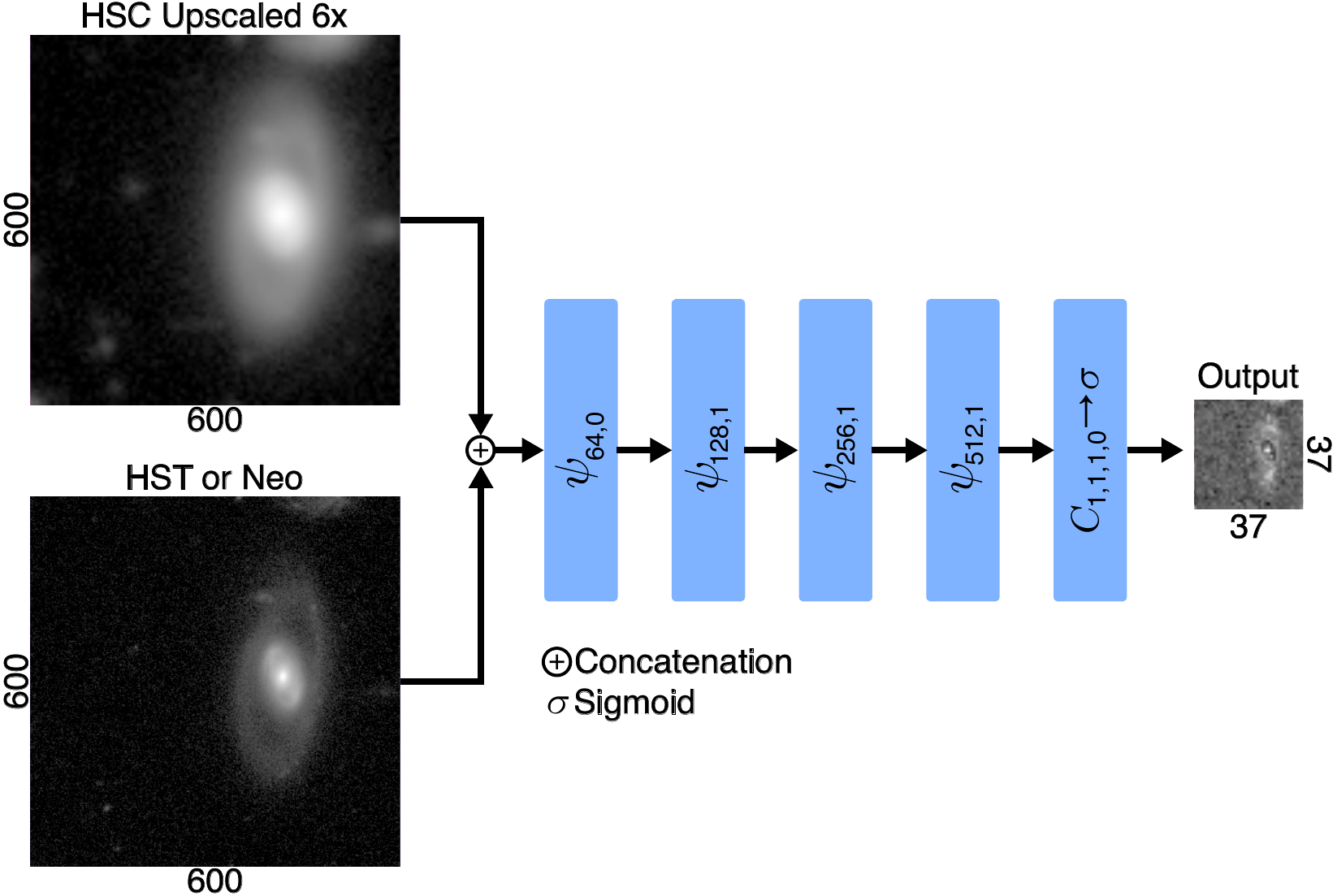}
\caption{Schematic of the PatchGAN discriminator network. The input tensors
        consist of two channels.
        The first channel is a nearest-neighbor up-sampling of the
        low-resolution ground-based
        image, and the second channel is the image for which the PatchGAN discriminator should
        provide a ``realness'' score. The image to score
        corresponds to either a cutout from the high-resolution data or the
        \modelname{}-generated image.
        The PatchGAN discriminator passes the two-channel tensor through
        four contracting convolutional layers ($\psi_{f,b}$), a fully-connected layer $C$,
        and a sigmoid
        function $\sigma$. The resulting output image is a $37^2$-pixel image where each
        pixel represents the ``realness'' of the corresponding receptive field.}
\label{fig:discriminator}
\end{figure*}

The adversarial approach of GANs leverages discriminator neural networks
to distinguish between real and generated model images, allowing the
training of the generator to produce model images that closely
resemble real data. For \modelname{}, we adopt the \textsc{PatchGAN} discriminator described in \citet{pix2pix}.
In contrast to a typical discriminator neural network, which classifies the entire image at one time, the \textsc{PatchGAN} discriminator classifies patches of input as real or synthetic. Classifying patches allows the discriminator to ``correct'' mistakes
made by the generator consistently over an entire image. Discriminators that classify whole images could be biased by very ``real'' or
``fake'' portions of the input image.
We present a diagrammatic description of the
discriminator used in this work in Figure \ref{fig:discriminator}.
When optimizing during training, we average the discriminator’s patch-level outputs to obtain a single scalar score that reflects how  ``real'' or ``fake'' the image appears when considering all patches collectively.

\subsection{Loss functions}
\label{sec:losses}

Loss functions allow for the quantitative assessment of how model-generated
output differs from desiderata.
During training, the parameters of a neural network are
optimized to minimize the loss function. \modelname{} and the PatchGAN
discriminator are trained simultaneously with different but complementary loss
functions.

For the PatchGAN discriminator, we use binary cross-entropy for the
loss function. Mathematically, this loss function can be written as
\begin{equation}
\mathcal{L}_{D}(\mathbf{x}, y) = y \log D(\mathbf{x}) + (1 - y) \log(1 - D(\mathbf{x})) \, ,
\label{discloss}
\end{equation}
\noindent
where $D$ is the PatchGAN discriminator,
and
$\mathbf{x} \in \mathbb{R}^{2 \times 600 \times 600}$ is a \modelname{} or
high-resolution data
image paired with a 6$\times$ up-sampled low-resolution ground-based image.
The quantity $y \in \{0, 1\}$ represents the
label associated with $\mathbf{x}$, corresponding to $1$ for
real high-resolution data images and $0$
for model images
generated by \modelname{}.

The composite loss function used to train \modelname{} is comprised of
four separate loss functions that target different areas of the
model performance.
The
first loss function, called the adversarial loss, is defined as
\begin{equation}
\mathcal{L}_{\textrm{Adv}}(\mathbf{z}) = -\log D(G(\mathbf{z}) \oplus N^{6\times}(\mathbf{z})) \, ,
\label{genloss_adversarial}
\end{equation}
\noindent
where $D$ is the PatchGAN discriminator, $G$ is \modelname{,} $\mathbf{z} \in
\mathbb{R}^{1 \times 100 \times 100}$ is a ground-based low-resolution
image, $\oplus$ represents concatenation
along the channel axis, and $N^{6\times}$ is a function that uses nearest-neighbor
interpolation to increase the resolution by a factor of 6 in width and height.
Reducing this loss corresponds with producing images that the PatchGAN
discriminator finds to be more likely to be real.

The second component of the composite
loss function, the reconstruction loss, is defined as
\begin{equation}
\mathcal{L}_{\textrm{Recon}}(\mathbf{x}, \mathbf{\hat{x}}) = n^{-1}\lVert \mathbf{x}-\mathbf{\hat{x}} \rVert_1  \, ,
 \label{genloss_reconstruction}
\end{equation}
\noindent
where $\mathbf{x}$ is a high-resolution
data image reshaped into a vector, $\mathbf{\hat{x}}$ is
a \modelname{}-generated image of the same area of the sky
reshaped into a vector, and $n=wh$ is the number of pixels in $\mathbf{x}$ and
$\mathbf{\hat{x}}$. Reducing this loss incentivizes the model
to generate pixel values close
to those in the high-resolution data image.

Segmentation loss is the third component of the composite
loss, and is defined as
\begin{equation}
    \mathcal{L}_{\textrm{Seg}}(\mathbf{x}, \mathbf{\hat{x}}, \mathbf{m})
    = \frac{\mathbf{m} \cdot (\mathbf{x}-\mathbf{\hat{x}})}
      {\lVert \mathbf{m} \rVert_1} \, ,
\end{equation}
\noindent
where $\mathbf{x}$ is a high-resolution data
image reshaped as a vector, $\mathbf{\hat{x}}$ is a
\modelname{} generated image of the same area of the sky
reshaped into a vector, $\mathbf{m}$ is a binary segmentation map for the same
area of the sky as $\mathbf{x}$, reshaped into a vector. The map $\mathbf{m}$ is
generated using \texttt{sep} \citep{sep} with the detection threshold (\texttt{thresh}) set to 3,
a minimum area requirement (\texttt{minarea}) of 5 pixels, the number of thresholds used in
deblending (\texttt{deblend\_nthresh}) is set to 32 levels and the minimum contrast ratio
(\texttt{deblend\_cont})  is set to 0.005. The resulting segmentation map is binarized so
that $m_i=1$ for pixels belonging to detected sources and $m_i=0$ otherwise.
The segmentation loss resembles the reconstruction loss
(Equation \ref{genloss_reconstruction}), except that the segmentation loss
only penalizes the pixels associated with a source.
Additionally, penalizing the loss for
pixels associated with sources improves the qualitative performance
of the model.

The last component of the loss function is the Visual Geometry
Group (VGG) perceptual loss
\citep{srgan}, $\mathcal{L}_{\textrm{VGG}}$. To compute this
loss, real and generated model images are provided to the
trained VGG convolutional
neural network \citep{simoyan_vgg2014}. The output features from
the VGG model layer are used to compute a mean squared error to
assess the perceptual realism of the generated model image.
The VGG loss is widely used in style transfer networks \citep[e.g.,][]{zhang2018a,hou2024a}, and
we leverage the loss in the \modelname{} model to improve the
perceived consistency between the output model images and the
high-resolution data.

\subsection{Training Procedure}
\label{sec:model:training}
\modelname{} is trained following the strategy used in \cite{pix2pix}. Training
runs were $2.5\times10^{6}$ steps with a batch size of 8. We use the Adam optimizer \citep{kingma2017adammethodstochasticoptimization} with learning rates of $2\times10^{-5}$ and $2\times10^{-4}$ for
the discriminator and generator, respectively. These model training choices were determined through
experimentation. While \modelname{} was trained for $2.5\times10^{6}$ steps, the
final model was selected based on its ability to recover out-of-sample
morphological parameters (Section~\ref{sec:morph_params}) and on visual
inspection of the generated images.

\section{Datasets}
\label{sec:dataset}

In this work, we super-resolve lower-resolution ground-based data to generate
model images that approximate higher-resolution space-based data.
Training the \modelname{} model to provide this
combination of super-resolution and style transfer
requires two datasets observed in the same region of the sky.
For the purpose of demonstrating the performance of
\modelname{}, we chose the well-known COSMOS field that has
undergone extensive observation by numerous facilities \citep{cosmos_hst}.
For our low-resolution\footnote{Here, we label the
ground-based data as ``low-resolution'' only to distinguish it from the
higher resolution space-based data. The HSC data is a remarkable dataset
of extremely high quality generated with sophisticated processing.
Our model benefits substantially from access to this high-quality public dataset.},
ground-based data, this study utilizes
observations from the HSC Telescope \citep{HSC_technical_2012, HSC_technical_2018}.
The higher-resolution, space-based data is
obtained from the Advanced Camera for Surveys (ACS) of HST.

\subsection{HSC Images}
\label{sec: hsc_images}
The ground-based observations targeted for super-resolution modeling
with \modelname{} come from the Public Data Release 2 of the
HSC \citep{HSC_dr2}, covering a portion of
the COSMOS field. Our study focuses on $i$-band images
($695\rm{nm}$–$845\rm{nm}$), with a limiting AB magnitude of
approximately $26.7$ for $5\sigma$ point sources and a pixel scale of $0.168$
arcseconds per pixel. The PSF equivalent full-width half maximum
of the images is $\sim1.3$ arcseconds \citep{HSC_dr2}.

\subsection{HST Images}
\label{sec: hst_images}
The CANDELS survey of the COSMOS field provides space-based observations for
training and testing. To approximately
match the HSC $i$-band filter, we select
HST F814W filter ($603\rm{nm}$-$1008\rm{nm}$) images.
The HST images serve as the high-resolution data input to our model training.
The HST F814W
PSF FWHM is approximately $0.09$
arcseconds, more than 10$\times$ sharper than the ground-based HSC images.
The F814W image depth
reaches $27.2$ AB magnitude for $5\sigma$ point source detection. The original
pixel scale ratio between the Advanced Camera for Surveys on HST and HSC is
approximately 5.8:1. Using the \textit{reproject} package in Python, the HST
image is resized to have a ratio of exactly 6:1 to enable
the input and output image sizes processed by
\modelname{} to correspond to an integer multiple.
The characteristics of the HSC and HST data are
summarized in Table \ref{table: characteristics}.

\begin{deluxetable}{lcc}[h]
\tablewidth{0pt}
\tablecaption{Summary of the HSC and HST Training Data}
\tablehead{\colhead{Property} & \colhead{HSC} & \colhead{HST}}
\startdata
Band (nm)                       & 695-845 & 603-1008 \\
Pixel scale (arcsec/pix)        & 0.168   & 0.03     \\
PSF FWHM (arcsec)               & 1.3     & 0.09     \\
Depth (AB magnitude)            & 26.7    & 27.2     \\
\enddata
\tablecomments{Summary of the HSC $i$-band and HST F814W
dataset characteristics, as reported by the original sources \citep{cosmos_hst, HSC_technical_2012,
HSC_technical_2018}. The depth is the $5\sigma$ detection
limit for point sources.\label{table: characteristics}}
\end{deluxetable}

\subsection{Training and Validation Data}
\label{sec:train_val_data}

The data used to train the model and validate its performance are
sampled from the overlap of
HSC and HST data in the COSMOS field, as described in Sections \ref{sec: hsc_images}
and \ref{sec: hst_images}.
The overlapping area is split such that 80\% of the area is
used for a training set and 20\% is used for validation data.
Specifically,
1.5 million cutouts are sampled from the training area
($\approx4953^2$ pixels in HST and $\approx2022^2$ pixels in HSC) and
approximately 400,000 samples from the validation area ($\approx2553^2$ pixels
in HST and $\approx1042^2$ pixels in HSC). While the
validation and training regions are exclusively separated,
within the validation and training regions random samples can overlap.
The partial overlap is intended to induce translational invariance.

\section{Results} \label{sec:results}

The performance of the trained \modelname{} neural network
was assessed in three ways: visual inspection, the
demonstrated recovery
of noise properties similar to the original high-resolution
images, and the quantitative improvement in recovering galaxy morphological parameters from the model super-resolution images
over measurements performed on ground-based images.

\subsection{Visual Inspection}
\label{sec:visual-inspection}

Visual inspection of  \modelname{} generated images is used to evaluate the
recovery of subtle features, evaluate results where no ground truth exists, and
identify poorly generated super-resolution images that the analysis pipeline
might overlook.

\begin{figure*}
\centering
\includegraphics[height=\textwidth]{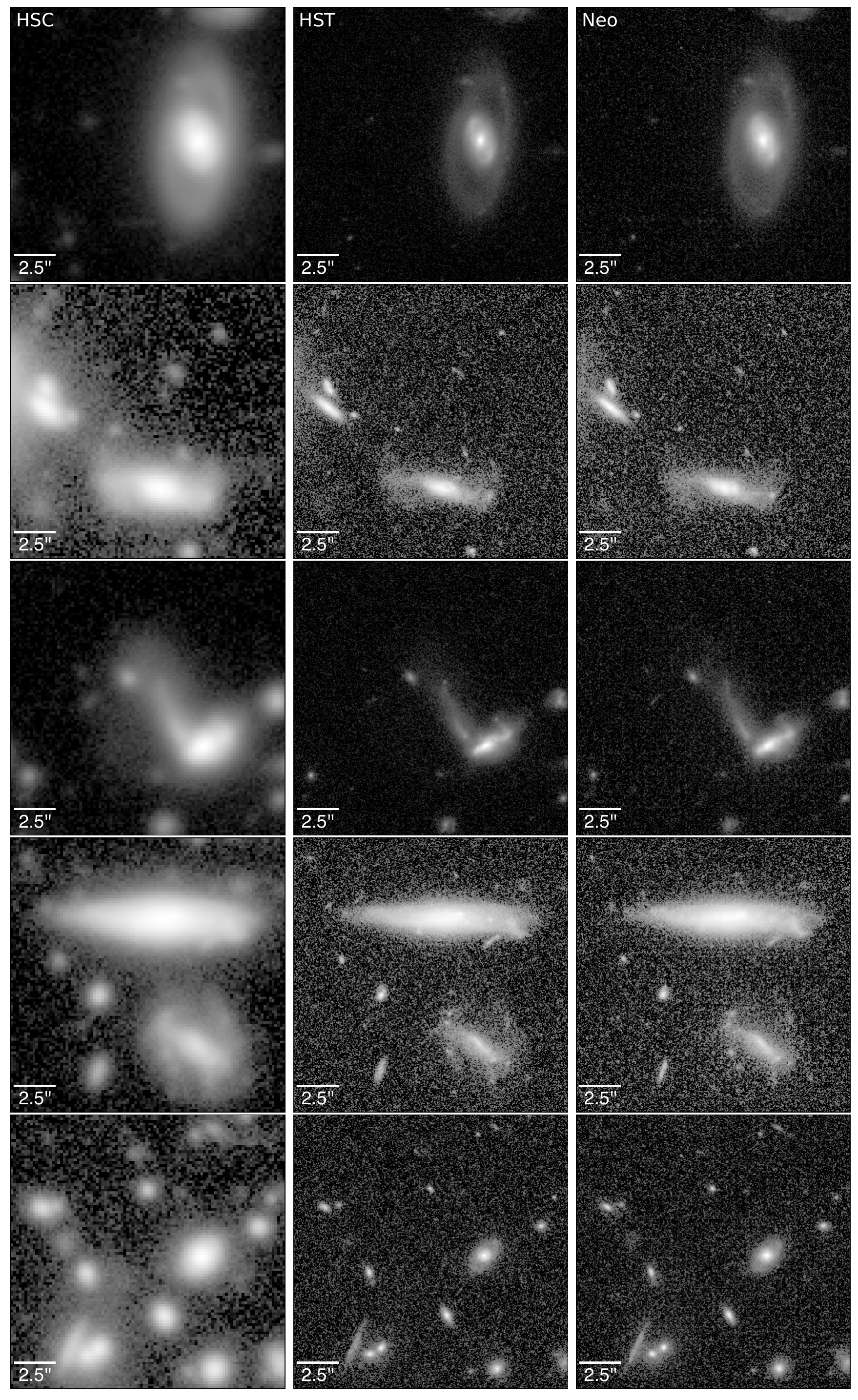}
\caption{Comparison of the super-resolution images generated by \modelname{}, and data from HSC and HST. The first column corresponds to the HSC images used as input to \modelname{}, the second column shows the generated super-resolution images when conditioned on the HSC image, and the last column presents the corresponding HST images. The input HSC $i$-band images are translated to target HST F814W images.
These cutouts show examples where significant visual improvements in orientation, effective radius, and axis ratio $q$ were found.
Fine structures such as rings (row 1), deblended objects (row 2, row 3, row 4, and row 5), and merging features (row 3) that were hardly visible at the native HSC resolution are present in the super-resolution images. Each of these images, pulled from the validation set, is stretched logarithmically to highlight faint surface brightness features.}
\label{fig:goodcutouts}
\end{figure*}

Figure \ref{fig:goodcutouts} shows five different HSC-HST-\modelname{} triplets, where the super-resolution images were generated by using the HSC image as an input. For reference, we include HST images, which serve as the ground truth. All images in Figure \ref{fig:goodcutouts} are from the validation set and were not used during training.

The selected validation
examples shown in Figure \ref{fig:goodcutouts} illustrate
the potential
of \modelname{} for producing high-fidelity super-resolution model
images of ground-based scenes.
Complex structures such as spiral arms, merging features,
or rings clearly present in the HST data
but hardly identifiable in HSC images are well recovered
by the \modelname{} model images. The images also demonstrate the ability
of \modelname{} to assist in deblending. For example, two distinct objects in the third row of Figure \ref{fig:goodcutouts} appear blended in the HSC data but are successfully deblended by \modelname{}, recovering the separation seen in the HST data. We emphasize that these images
are drawn from the validation data set, and represent scenes
from the HST and HSC data that were not used to train our model.

\begin{figure*}
\centering
\includegraphics[width=\linewidth]{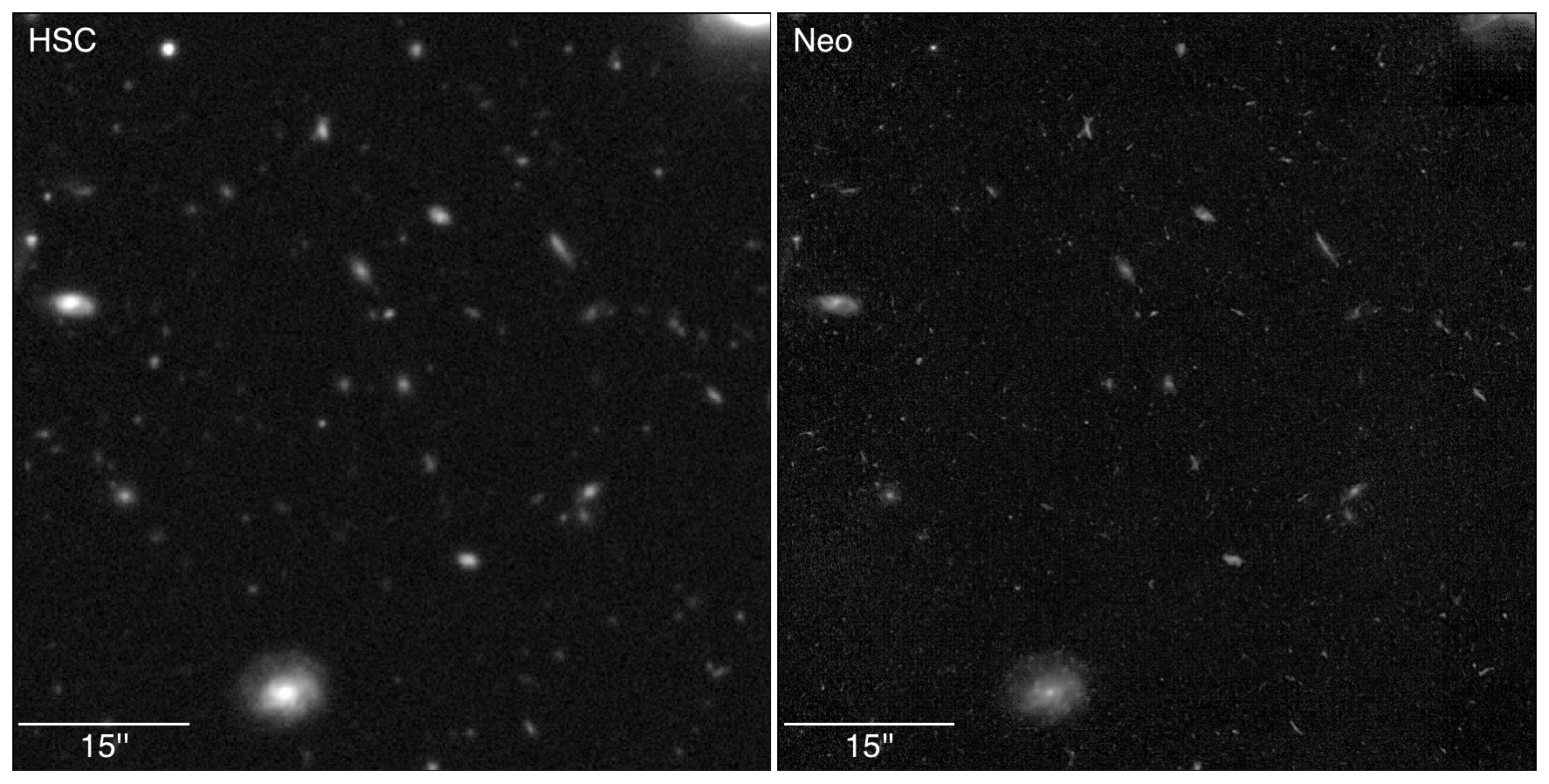}
\caption{Super-resolved \modelname{} image (right panel) of an HSC field (left panel)
         never seen by HST (DEEP2-3). The input HSC images are in I-band. This figure
         illustrates that \modelname{} can
         produce realistic results for areas of the sky never seen during
         training. The field shown is $15\times15\,\rm{arcseconds}$, i.e.
         $400^2$ pixels for the HSC image and $2400^2$ pixels
         for the \modelname{} image.}
\label{fig:sr-mosaic}
\end{figure*}

\modelname{} can also generate super-resolution mosaics, which
we demonstrate using data from the DEEP2-3 survey and
PDR3 (Public Data Release 3).
\modelname{} was trained on overlapping HSC-HST areas of the COSMOS field,
and then applied to the DEEP2-3 survey data that is completely
independent of COSMOS.
By revealing a high-quality super-resolution version of the DEEP2-3
survey data, Figure \ref{fig:sr-mosaic} shows the ability of \modelname{} to generalize to other parts of the sky beyond our training sample.
The ability to perform well in new areas of the sky is crucial for
the general applicability of \modelname{}, as we anticipate
applying \modelname{} on datasets beyond the COSMOS HST/HSC pairing.
The original ground-based low-resolution data shown
in Figure \ref{fig:sr-mosaic} has dimensions of $400\times400$, resulting
in a $2400\times2400$ super-resolution image generated by \modelname{}.
Here, the \textsc{Morpheus} parallel stitching algorithm \citep{hausen2020} was applied with a stride of 20 and a window size of $100\times100$, producing a super-resolution mosaic constructed from 256 HSC cutouts super-resolved by \modelname{}. In Figure \ref{fig:sr-mosaic}, the qualitative
improvement in the super-resolution image versus HSC is clearly visible, and since no corresponding HST data exists, these images provide a model prediction
for what a high-resolution view of this area might resemble. We expect to
apply \modelname{} in many other areas of the sky without HST or comparable
space-based images, to provide high-resolution model images that reliably
recover the expected resolved object properties. We turn to demonstrating
that capability quantitatively below.

\subsubsection{Color Images}
\label{sec:color-images}
\begin{figure*}
\centering
\includegraphics[width=\linewidth]{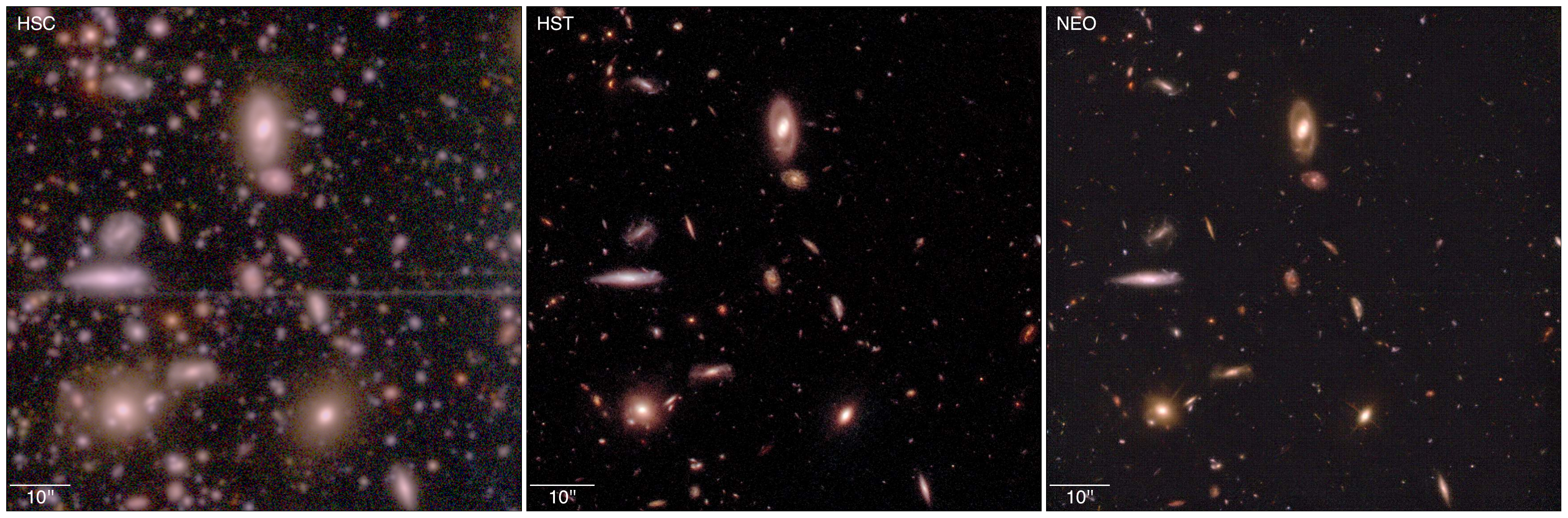}
\caption{False color images comparing HSC, HST, and  \modelname{} results for the same field
drawn from the validation dataset.
For HSC, the $i$, $r$, and $g$ filters are mapped to the red, green, and blue channels, respectively. The \modelname{} images are generated from the input HSC $i$, $r$, and $g$
images and use the same RGB mapping as HSC.
For HST, the WFC3
F160W filter data are mapped to the red channel, the F814W filter data are mapped to the green channel, and the F606W filter data are mapped to the blue channel. The F160W data are reprojected to match the pixel scale of the F606W and F814W images. The input HSC mosaic is $536\times536$ pixels and the HST and \modelname{} mosaics are $3216\times3216$ pixels.}
\label{fig:color-images}
\end{figure*}

As an additional qualitative test, we use \modelname{} to generate
RGB images from multiband HSC images.
Figure \ref{fig:color-images} shows a false color
RGB image generated from HSC $i$, $r$, and $g$-band
images (left panel) with $536\times536$ pixels. After
processing the images with the trained
\modelname{} model, super-resolution images
with $3216\times3216$ pixels are generated and an
RGB image constructed with the same color
scaling as the HSC data (right panel).
For comparison, we show
an F160W/F814W/F606W HST RGB\footnote{The HST WFC3 F160W filter image has a different pixel scale than
the HST ACS data, and is reprojected to match the pixel scale of the F814W and F606W images.}
image (center) for
visual comparison with the resolved structures
seen in the \modelname{} model images.
While \modelname{} was only trained on HST $i$-band, the visual
performance of \modelname{} generalizes to other filter data. The performance across different
input filter images indicates
that \modelname{} remains
robust to changes in image characteristics over different bands.

\subsection{Noise Properties}\label{sec:noise_char} 
\begin{figure}
\centering
\includegraphics[width=\linewidth]{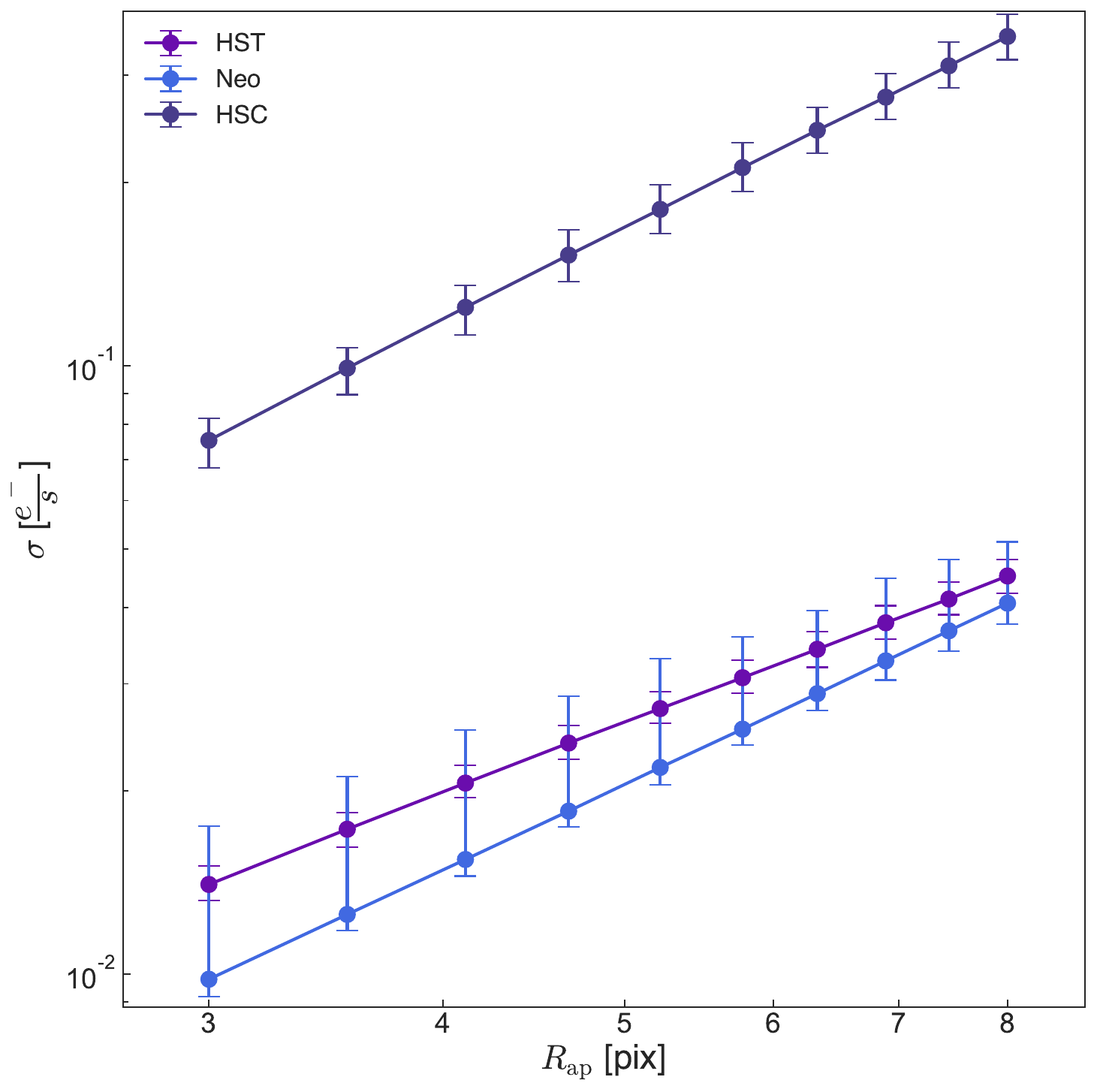}
\caption{Comparison of the noise properties of HST, \modelname{}, and HSC.
          $R_{\rm{ap}}$ is the aperture. $\sigma$ is the standard deviation of the sum of the
pixel values calculated by randomly placing $200$
circular apertures with radius $R_{\rm{ap}}$ on the images.}
\label{fig:noiseplot}
\end{figure}

The noise properties of the generated \modelname{} images are examined and compared to those of HST and HSC. This analysis verifies whether \modelname{} can approximate the background statistics and depth of the
HST images when generating super-resolution model
images from HSC data. First, a super-resolution mosaic is created from individual HSC cutouts by generating a super-resolution cutout for each image and then combining them into a large-format image
using the \textsc{Morpheus} stitching
algorithm. The noise properties of the super-resolution mosaic are then compared to the equivalent field in HST.
Figure \ref{fig:noiseplot} shows the measurements of the noise properties for the three mosaics (HSC, HST,
and \modelname{}).
To estimate the depth, sources in the field have been detected and masked via \texttt{photutils} \citep{photutils}, with sigma-clipping applied to the resulting sky pixel
distribution. To measure how the image RMS uncertainties
scale with aperture size,
a power-law model is fit
for HST, \modelname{}, and HSC data independently as
\begin{equation}
    \log \sigma(R_{\rm{ap}}) = a\cdot\log R_{\rm{ap}} + b
\label{eq:noisepowerlaw}
\end{equation}
\noindent
where $\sigma$ is the standard deviation of image counts within an aperture, $ R_{\rm{ap}}$ is
the aperture radius measured in pixels, $a$ is the slope of the relationship,
and $b$ is the single-pixel limit of the RMS.
The fit results are in Table~\ref{table: noiseprop}.

\begin{deluxetable}{lcc}[h]
\tablewidth{\textwidth}
\tablecaption{Scaling of Noise with Aperture Size}
\tablehead{\colhead{Image} & \colhead{Index $a$} & \colhead{Interecpt $b$}}
\startdata
HST             & 1.18  & -5.56 \\
\modelname{}    & 1.45  & -6.22     \\
HSC             & 1.55  & -4.28    \\
\enddata
\tablecomments{Scaling of the rms noise in HST, HSC, and \modelname{} as a function
of aperture size. The Table reports results of fitting Equation \ref{eq:noisepowerlaw} for measurements of the rms of noise measured in random circular apertures of different
sizes, for the HST, \modelname{}, and HSC images.
\label{table: noiseprop}}
\end{deluxetable}

By inspection, it appears \modelname{}
recovers the depth of HST, but with increased
pixel-level covariance in the noise as
quantified by the power-law index $a$. Given that
the super-resolution model has to extrapolate below
the native pixel scale of the ground-based imaging,
additional pixel covariance relative to HST
should not be surprising.
Other differences in the values shown in
Table \ref{table: noiseprop} could arise
from, for example, the details of the
method used by \modelname{} to super-resolve the background, how negative noise values are injected
during the Crop-InvScale step of the model,
and potentially our mosaicking technique.
Given the
complexity of these factors and the fact that galaxy morphological parameter fitting in the \modelname{}
model
images matches well the results on HST data (as discussed in Section
\ref{sec:morph_params}), we leave a more detailed
examination of the noise properties to future work.

\subsection{Morphological Properties}
\label{sec:morph_params}

The results presented in
Section \ref{sec:visual-inspection} suggest that the
\modelname{} model can generate super-resolution
images that resemble high-resolution, space-based
data from lower resolution, ground-based data.
We can perform quantitative analyses to determine
whether the morphological properties of objects
in the super-resolution model image match those
expected from the high-resolution training dataset,
and whether the super-resolution image provides
a means for accurately determining the properties
of objects in low-resolution imaging as if they
were seen by an actual high-resolution facility.
Here, we measure a range of morphological
properties of objects in the HSC, \modelname{},
and HST images and show that \modelname{}
can indeed recover the HST-determined
properties of HSC-observed objects using
low-resolution ground-based imaging of
galaxies beyond its training set.

To measure the morphological properties of the HST, HSC and \modelname{} images,
a pipeline was developed based on \texttt{photutils} where the HST segmentation
maps are considered as ground truth for source locations. A custom pipeline was
used rather than an existing catalog to ensure consistency between measurement
methods over all three image sources. The pipeline overlays the HST segmentation
map onto the \modelname{} image for measuring morphological parameters in
\modelname{} images. In the case of HSC images, the procedure involves
re-projecting the HST segmentation map onto the HSC pixel scale, thereby
creating a lower-resolution segmentation map conditioned on the HST pixel
locations.

For this quantitative analysis, we measure the
effective radius ($R_e$), ratio of the radii enclosing respectively \(75\%\)
and \(25\%\) of the total flux $C_{75/25}$,
 on-sky orientation, full-width half maximum (FWHM), and $q = b/a$, the
ratio of the semi-minor axis to the semi-major axis.
To quantify the relative agreement between the morphological
parameters measured for galaxies in the three imaging
datasets, several metrics are defined.
For the set of morphological parameters $p \in [R_{\mathrm{e}}, \rm{FWHM}, C_{75/25}]$, we calculate a fractional relative bias between
the HSC or \modelname{} measurements and the measurements performed on HST data. We define
a relative bias
\begin{equation}
\label{eq:relative_bias}
    \mathcal{\tilde{B}}_{p} = \dfrac{\mathcal{M}_{p, X} - \mathcal{M}_{p, {\rm{HST}}}}{\mathcal{M}_{p, {\rm{HST}}}}\,,
\end{equation}
\noindent
where $\mathcal{M}_{p, \rm{HST}}$ are the
measurements on the HST data and $\mathcal{M}_{p, X}$
represents a measure of the parameter $p$ on the
images of dataset $X$, with $X\in[$\modelname{},HSC$]$.
Similarly, we define the absolute bias $\mathcal{B}_{p}$ for the axis-ratio $q$
as
\begin{equation}
\label{eq:abs_bias}
    \mathcal{B}_{p} = \mathcal{M}_{p, X} - \mathcal{M}_{p, {\rm{HST}}}\,
\end{equation}
\noindent where $\mathcal{M}_{p, \rm{HST}}$ and
$\mathcal{M}_{p, X}$ are the same as in Equation \ref{eq:relative_bias}.

The on-sky orientations, or position angles, of galaxies in
the HST and HSC or \modelname{} data are
compared using 1 minus the absolute
value of the cosine similarity of the unit vectors with the same angle as
\begin{equation}
    \label{eq: sim diff}
    \mathcal{S} = 1 - \left|S_c(\hat{\mathbf{u}}_{\modelname{}}, \hat{\mathbf{u}}_{\mathrm{HST}})\right|
\end{equation}
\noindent
where $S_c$ is the cosine similarity.

\begin{deluxetable}{lccc}[h]
\tablewidth{0.5\textwidth}
\tablecaption{Custom Pipeline Validation}
\tablehead{\colhead{Pipeline} & \colhead{$R_e$} & \colhead{$q$} & \colhead{Orientation}}
\startdata
Custom  & \textbf{-5.2$\pm$ 0.5$\boldsymbol{\times 10^{-3}}$ } & \textbf{0.30 $\pm$ 0.81} & \textbf{0.99 $\pm$ 0.13} \\
SExtractor & 3.1$\pm$ 0.5$\times 10^{-2}$ & 0.48 $\pm$ 1.02 & 0.99 $\pm$ 0.15 \\
\enddata
\tablecomments{Comparison of the custom pipeline used in this analysis and an SExtractor-based pipeline, validating the performance of our custom pipeline for morphological analysis.
Measurements of effective radius $R_e$, axis ratio $q$, and on-sky orientation were
performed for galaxies in the DREAM simulated images, and then compared with the
input DREAM catalog values. The Table reports the measurement bias (Equation \ref{eq:relative_bias}) for SExtractor and the custom pipeline relative to the DREAM
input catalog values. Bold indicates better performance than the alternative pipeline.
\label{table:dream}}
\end{deluxetable}

The accuracy of the morphological measurement pipeline is validated by processing simulated images generated
from the DREAM catalog and simulated image suite \citep{Drakos_2022}. SExtractor \citep{sextractor} is also used as a baseline comparison due to its pervasive adoption.
Equation \ref{eq:relative_bias} is used to measure the bias between the DREAM catalog and both the custom pipeline and SExtractor. The results are summarized in Table \ref{table:dream}, showing that the custom pipeline performs better than SExtractor for the relevant morphological parameters provided in the DREAM catalog. Validation is performed with the morphological parameters $R_e$, $q$, and orientation as they are all that is directly comparable in the DREAM catalog. Note that while $q\in[0,1]$, dispersions greater than 1 can occur because of how Equation~\ref{eq:relative_bias} is defined.

Extending the analysis to the real data,
for each of the comparison metrics we compute the
HSC, \modelname{} and HST properties for each galaxy.
We measure
morphological parameters from cutouts of size
$100\times100$ for HSC cutouts and $600\times600$ for
HST and \modelname{}.
Table \ref{table:results}
provides a summary of the results, reporting
the median of these
metrics computed over the whole test set.
Results for measuring the morphological
parameters where the cutouts have been integrated into a mosaic are also provided.
As Table
\ref{table:results} shows, the measured morphological values from the \modelname{}
generated images are significantly closer to the measured values for the
HST images than those measured from the HSC images for both single cutouts and mosaics.

Figure \ref{fig:marginals} shows the average measurement
of the morphological parameters (clockwise from upper left) effective radius $R_e$, concentration $C_{75/25}$,
FWHM, and axis ratio $q$ in bins of HST F814W magnitude.
The distributions of the morphological parameter metrics
are shown quantitatively in Figures \ref{fig:re}--\ref{fig:orientation}.
Each of these figures follows the same four-panel structure.
For each morphological parameter metric, the first and third panels show 2D-histograms of the metric from HSC (left) and \modelname{} (right) as a function of the magnitude of the same object as measured in the HST cutout. The second panel of each figure overlays the HSC (gray) and \modelname{} (blue) distributions. The fourth,
rightmost panel shows the marginal distribution of the
morphological metric.
Note that the axis ranges are clipped for better visibility,
such that outliers are excluded. These outliers
represent very few objects out of the 18,783:
only 8 for $C_{75/25}$, 43 for orientation, 62
for FWHM, 74 for $R_e$, and 9 for $q$ are excluded by
these cuts.

Since both training and evaluation use randomly drawn, possibly overlapping cutouts from their respective (non-overlapping) regions, objects are likely to be measured more than once under different crops. This design reflects the intended augmentation (translation invariance) and enlarges the effective sample without changing the qualitative comparisons between HSC, \modelname{}, and HST. These cutout-level metrics are reported below. Representative failure modes and their practical impact are summarized in Appendix~\ref{Appendix A}. We next quantify improvements for each morphological parameter.

\begin{deluxetable*}{llccccc}[h]
\tablewidth{0pt}
\tablecaption{Morphological Parameter Performance}
\tablehead{\multicolumn{2}{c}{Image} & \colhead{$R_e$} & \colhead{FWHM} & \colhead{$q$} & \colhead{$C_{75/25}$} & \colhead{Orientation}}
\startdata
\multirow{2}{*}{\modelname{}} & single & \textbf{0.04 $\pm$ 0.30} & \textbf{-0.02 $\pm$ 0.06} & \textbf{0.03 $\pm$ 0.04} & \textbf{8.70$\pm$ 0.11$\boldsymbol{\times 10^{-3}}$ } & \textbf{5.60$\pm$ 0.10$\boldsymbol{\times 10^{-4}}$ } \\
                              & mosaic & \textbf{0.11 $\pm$ 0.15} & \textbf{0.04 $\pm$ 0.05} & \textbf{0.03 $\pm$ 0.04} & \textbf{0.09 $\pm$ 0.11} & \textbf{1.54 $\pm$ 0.06$\boldsymbol{\times 10^{-3}}$} \\
\multirow{2}{*}{HSC}          & single & 0.72 $\pm$ 0.18          & 0.82 $\pm$ 0.05 & 0.11 $\pm$ 0.14 & -0.20 $\pm$ 0.11 & 6.6$\pm$ 0.17 $\times 10^{-3}$ \\
                              & mosaic & 0.65 $\pm$ 0.98          & 0.13 $\pm$ 0.24 & 0.11 $\pm$ 0.14 & 0.11 $\pm$ 0.11 & 0.01 $\pm$ 0.18 \\
\enddata
\tablecomments{Comparative results for \modelname{} and HSC across key morphological
         parameters using equations \ref{eq:relative_bias}--\ref{eq: sim diff}. Each column represents a different parameter: $R_e$ is the
         effective radius, FWHM is the full width at half maximum, $q$
         represents the ellipticity, $C_{75/25}$ is the flux concentration, and
         orientation measures the object's angle of orientation in space. The
         table compares the performance and dispersion for \modelname{} and HSC for both
         single cutouts and mosaic measurements.\label{table:results}}
\end{deluxetable*}

\subsubsection{Effective Radius}
\begin{figure*}
\centering
\includegraphics[width=\linewidth]{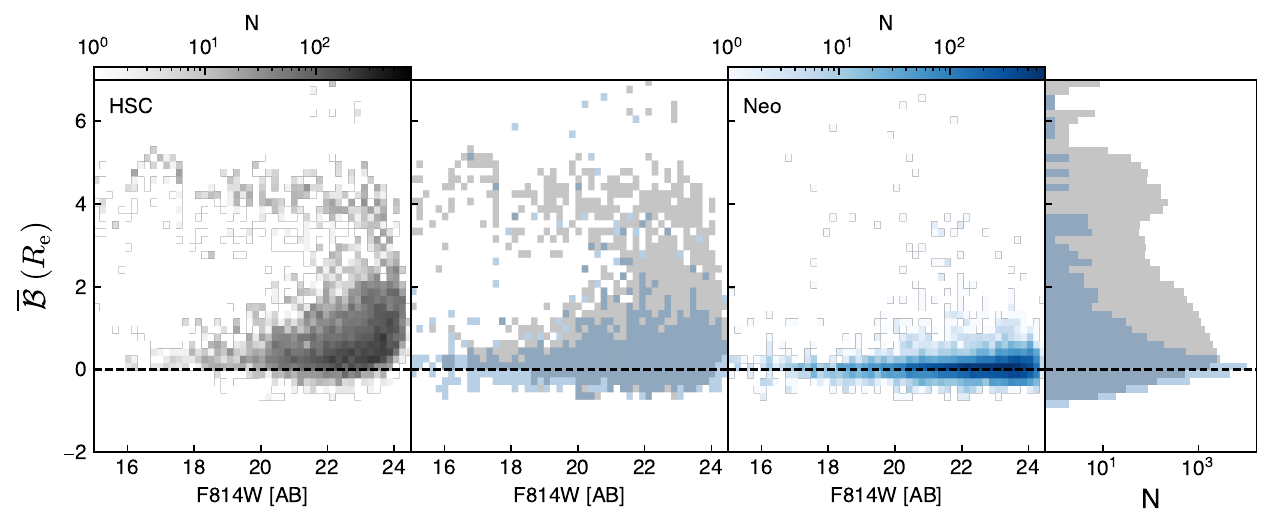}
\caption{Comparison of the fit of the half-light radius, $R_{e}$, for the same
         galaxies on different images. The dependent-axis represents the relative bias $\bar{\mathcal{B}}$
         (eq.~\ref{eq:relative_bias}) while the independent-axis shows the HST magnitude of the
         fitted objects. The first and third panels show the 2D histograms of the
         results for $\bar{\mathcal{B}}$ vs. $R_{e}$ measured for
         the HSC (first) images and \modelname{}'s super-resolved images
         (third). The second panel overlays the two 2D histograms for a more direct
         comparison. Finally, the right-most panel shows the projected 1D
         histogram of the bias, for all magnitudes. For better visualization,
         the $\bar{\mathcal{B}}$-axis excludes objects more than 5$\sigma$ from the mean.
         HSC overestimates $R_{e}$ across all magnitudes, becoming more pronounced for fainter objects. In contrast, \modelname{} improves $R_{e}$ measurements across the entire magnitude range.}
\label{fig:re}
\end{figure*}

The mean relative bias \(\mathcal{\tilde{B}}_{R_{e}}\) in the effective radius  morphological parameter $R_e$
measured from \modelname{} images is $0.04 \pm 0.30$ compared with the same galaxies
in HST images, significantly improving upon the
bias $0.72 \pm 0.18$ measured from the HSC data (Table \ref{table:results}). In mosaics, the relative bias is $0.11 \pm 0.15$, which similarly outperforms the HSC mosaic bias of $0.65 \pm 0.98$  (Table \ref{table:results}).
Figure~\ref{fig:marginals} and
 Figure~\ref{fig:re}
show that \modelname{} recovers $R_{e}$ across all magnitudes with only a slight overestimation for faint objects. Measurements on the HSC data
systematically overestimate $R_{e}$ across all magnitudes, becoming more pronounced for fainter galaxies.
This analysis demonstrates that the half-light
radii $R_{e}$ of galaxies measured in the \modelname{} model
images reliably recover the sizes of galaxies
in HST images of the same area of the sky, over a
wide range of galaxy brightnesses.

\subsubsection{Full-Width Half-Maximum}
\begin{figure*}
\centering
\includegraphics[width=\linewidth]{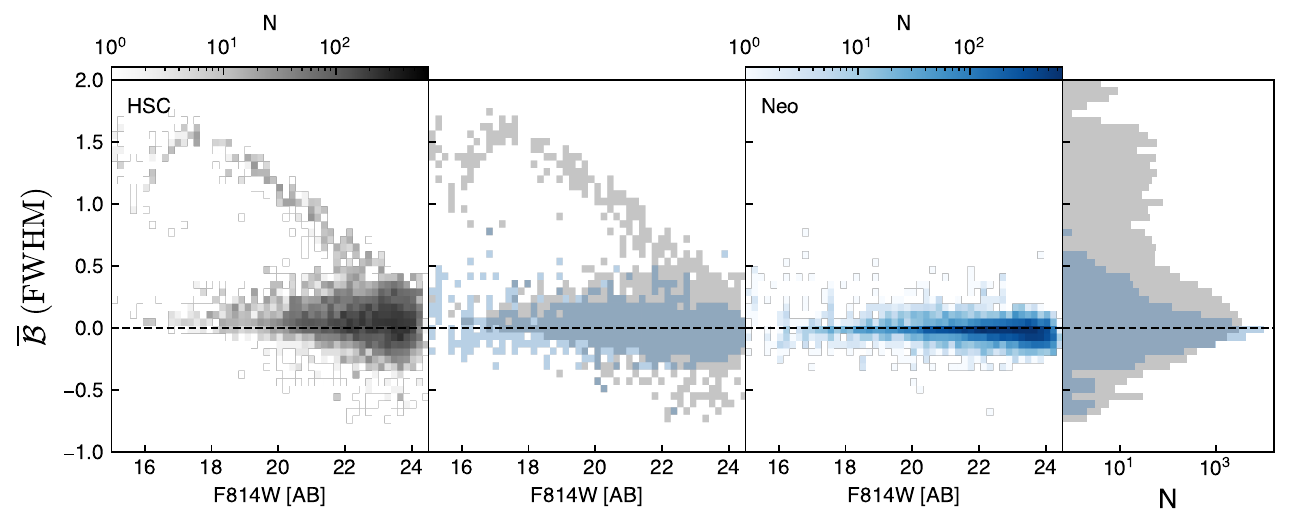}
\caption{Comparison of the fitted full-width-half-maximum, FWHM, with F814W magnitude
         for the
         same galaxies on different images. The format follows Figure \ref{fig:re}. The HSC images consistently overestimate the FWHM of objects, particularly for bright galaxies, and exhibits a much broader scatter. The \modelname{} model recovers the FWHM of objects across the entire distribution of magnitude, with a tighter spread and no clear bias to overestimation or underestimation.}
\label{fig:fwhm}
\end{figure*}

The mean relative bias
in the light distribution full-width half-max (FWHM)
measured from \modelname{} images
is $\mathcal{\tilde{B}}_{\rm{FWHM}} = -0.02 \pm 0.06$, which represents
a substantial improvement over the
bias of $0.82 \pm 0.05$ measured from the HSC images
(Table \ref{table:results}). For mosaics, \modelname{} achieves a relative bias of $0.04 \pm 0.05$, again performing better than the HSC mosaic bias of $0.13 \pm 0.24$  (Table \ref{table:results}).
Figure~\ref{fig:marginals} and
Figure~\ref{fig:fwhm}
 demonstrate that \modelname{} generates images
that recover the FWHM values nearly identical to HST across all magnitudes, with slight deviations only for faint objects. In contrast, measurements on the
HSC images consistently overestimate the
FWHM, particularly for bright galaxies, and exhibit a much wider scatter. This analysis
shows  \modelname{} delivers images that
accurately recover the light distribution
FWHM measurements across galaxy brightness and size.

\subsubsection{Concentration}
\begin{figure*}
\centering
\includegraphics[width=\linewidth]{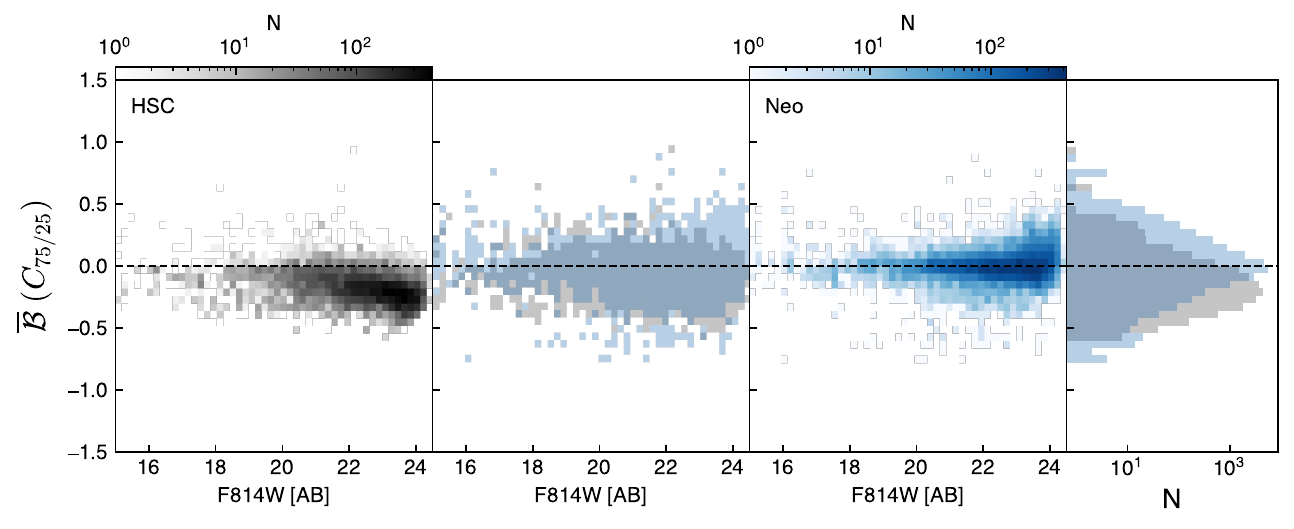}
\caption{Comparison of the fitted concentration parameter $C_{75/25}$ as a function
         of F814W magnitude for
         the same galaxies as measured on the HSC and corresponding \modelname{} super-resolution
         images. HSC  underestimates $C_{75/25}$ at all magnitudes, resulting in a clear systematic negative bias that worsens for higher-magnitude objects. In contrast, \modelname{} recovers galaxy concentrations across a variety of brightness and structure.}
\label{fig:r75_25}
\end{figure*}

For the light concentration of objects,
the mean relative bias \(\mathcal{\tilde{B}}_{C_{75/25}}\) measured from \modelname{} images is $8.7\times10^{-3} \pm 0.11$, while the bias measured from HSC images is $-0.20 \pm 0.11$ (see Table \ref{table:results}). For mosaics, the relative bias is $0.09 \pm 0.11$, again an improvement compared to the HSC mosaic bias of $0.11 \pm 0.11$  (Table \ref{table:results}).
Figure~\ref{fig:marginals} and
Figure~\ref{fig:r75_25}  show that \modelname{} recovers $C_{75/25}$ values closely matching HST across all magnitudes, with only slightly increased scatter at faint magnitudes. In contrast, the concentration measurements
on HSC images consistently underestimate
$C_{75/25}$ at all magnitudes, resulting in a clear systematic negative bias.

\subsubsection{Axis Ratio}
\begin{figure*}
\centering
\includegraphics[width=\linewidth]{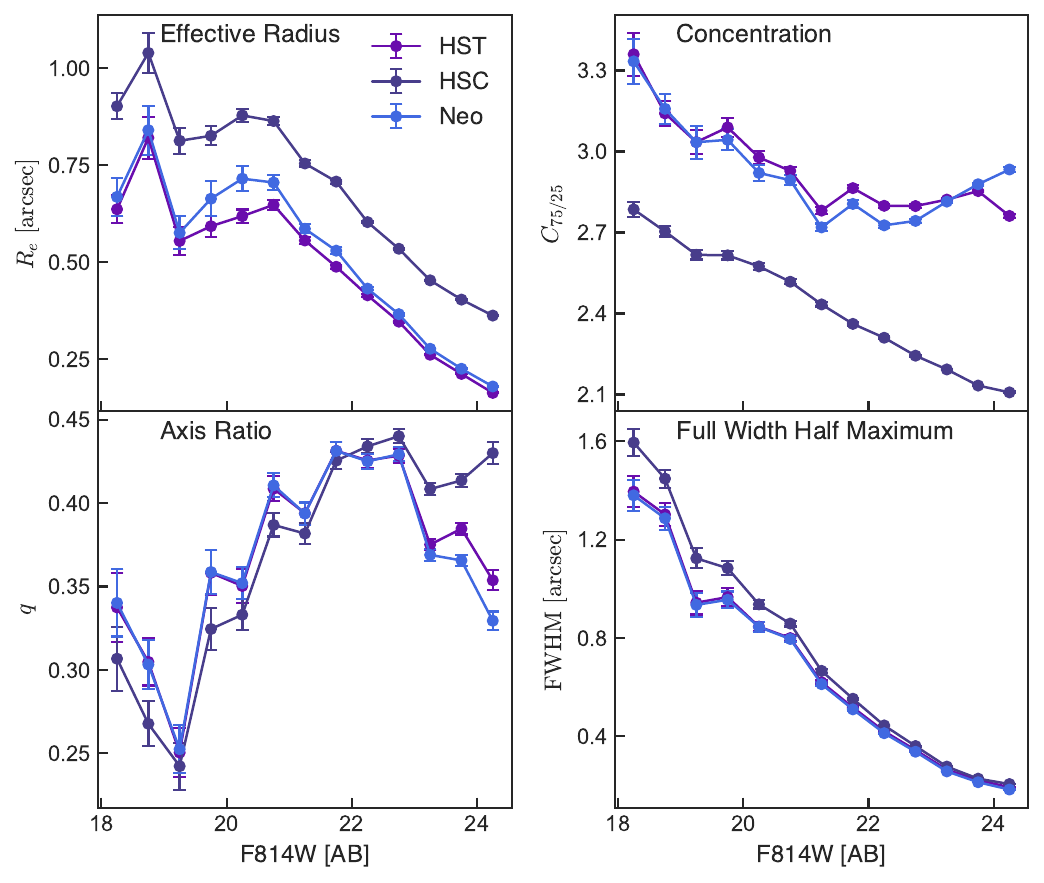}
\caption{Comparison of galaxy morphological parameters measured for HSC (indigo), HST (purple), and \modelname{} (blue) across galaxy brightness measured in F814W magnitude. Each subplot shows median values computed in incremental 0.5 magnitude bins, with error bars representing the standard error of the mean within each bin. We see that HSC systematically overestimates $R_{e}$ and FWHM, and consistently underestimates the flux concentration $C_{75/25}$. Additionally, HSC ellipticity measurements $q$ show a magnitude-dependent bias, underestimating ellipticity at low magnitudes and overestimating at higher magnitudes. In contrast, \modelname{} closely reconstructs all HST measurements, significantly reducing systematic biases and scatter, while demonstrating accurate recovery across the full range of galaxy magnitudes and morphologies.}
\label{fig:marginals}
\end{figure*}

\begin{figure*}
\centering
\includegraphics[width=\linewidth]{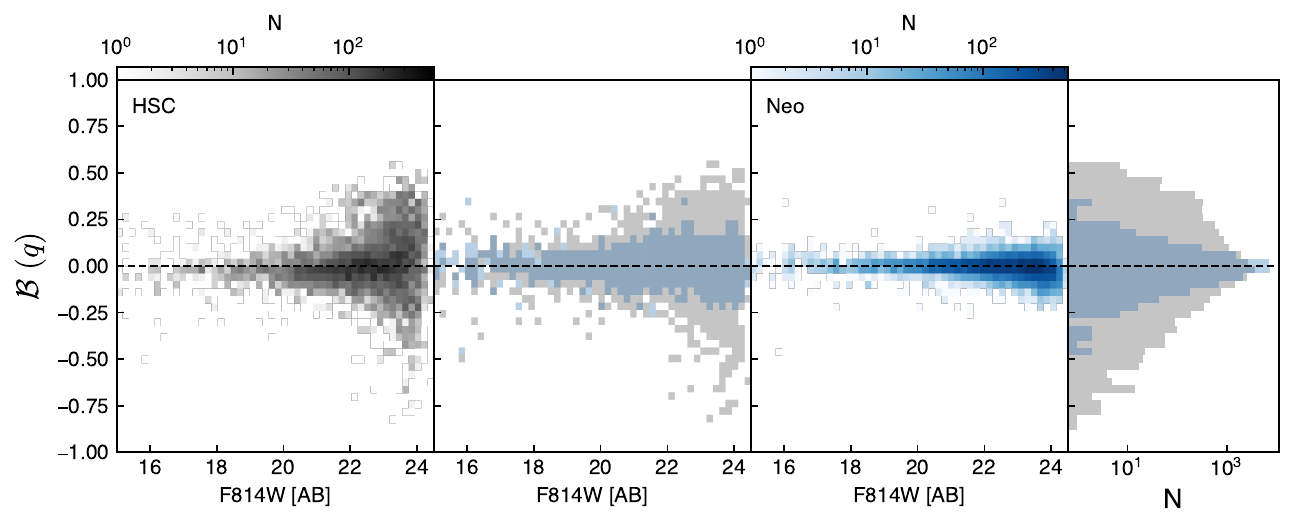}
\caption{Comparison of the fitted axis-ratio, $q$, as a function of F814W magnitude for the
         same galaxies
         in HSC and \modelname{} images. The format follows Figure \ref{fig:re}. HSC slightly underestimates  \(q\) at brighter magnitudes and then overestimates $q$ for faint galaxies, albeit with large scatter. In contrast, \modelname{} recovers shape measurements across the full range of galaxy morphologies and magnitudes. The \modelname{} model improves axis-ratio measurements across the entire magnitude range, with the largest improvements found for brighter objects.}
\label{fig:q}
\end{figure*}

The ratio of semi-minor to semi-major axes
of galaxies measured from the \modelname{} images
has a mean absolute bias $\mathcal{B}_{q} = 0.03 \pm 0.04$, compared to a
bias of $0.11 \pm 0.14$ measured from the HSC data for the
same objects (Table \ref{table:results}). When measuring on mosaics, \modelname{} yields a bias of $0.03 \pm 0.04$, comparable to its single-image performance and better than the HSC mosaic bias of $0.11 \pm 0.14$  (Table \ref{table:results}).
Figure~\ref{fig:marginals} and Figure~\ref{fig:q} show that \modelname{} recovers the \(q\) measured in the HST data
consistently across all magnitudes, with only minor scatter at faint magnitudes. Measurements
performed on the HSC images underestimate
\(q\) (galaxies appear more flattened) at bright magnitudes and overestimate \(q\) (galaxies look rounder) for faint objects.
Measurements on the \modelname{} images recover
HST
shape measurements across the full range of galaxy morphologies and brightness levels,
with substantially less scatter than measurements on the HSC images.

\subsubsection{Orientation}

\begin{figure*}
\centering
\includegraphics[width=\linewidth]{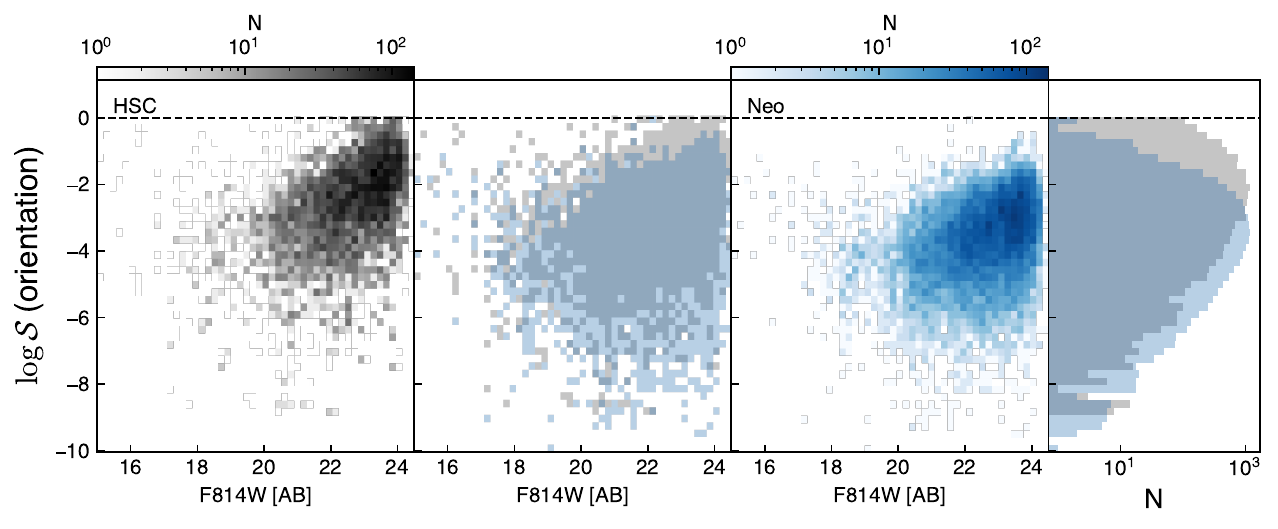}
\caption{Comparison of the fitted orientation parameter $\mathcal{S}$ for the same
         galaxies measured in HSC and \modelname{} images, as a function of F814W magnitude.
         The format follows Figure \ref{fig:re}.
         Galaxies in the HSC show less alignment with the orientation measured from HST images, with degrading agreement for objects with fainter magnitudes. The galaxies in the
         \modelname{} images show better overall alignment with HST-measured orientations across
         the whole magnitude range.
         }
\label{fig:orientation}
\end{figure*}

The mean similarity measure \(\mathcal{S}\) for orientation measured from \modelname{} images
is $5.6\times10^{-4} \pm 0.10$,
compared to the value of $6.6\times10^{-3} \pm 0.17$
measured from the HSC images
(Table \ref{table:results}). For mosaic cutouts, the relative bias is $1.54 \times 10^{-3} \pm 0.06$, an improvement over HSC mosaics, which show a bias of $0.01 \pm 0.18$  (Table \ref{table:results}).
Figure~\ref{fig:orientation} demonstrates
that \modelname{} recovers galaxy orientations measured by HST across the full magnitude range, with negligible scatter even for faint galaxies. Measurements on the ground-based
HSC show greater deviations, especially at faint magnitudes, resulting in less accurate orientation recovery.
Measurements on the \modelname{} images
recover accurate
galaxy orientations across all magnitudes present in our
validation dataset.

\subsection{Point Spread Function}
\begin{figure*}
\centering
\includegraphics[width=\linewidth]{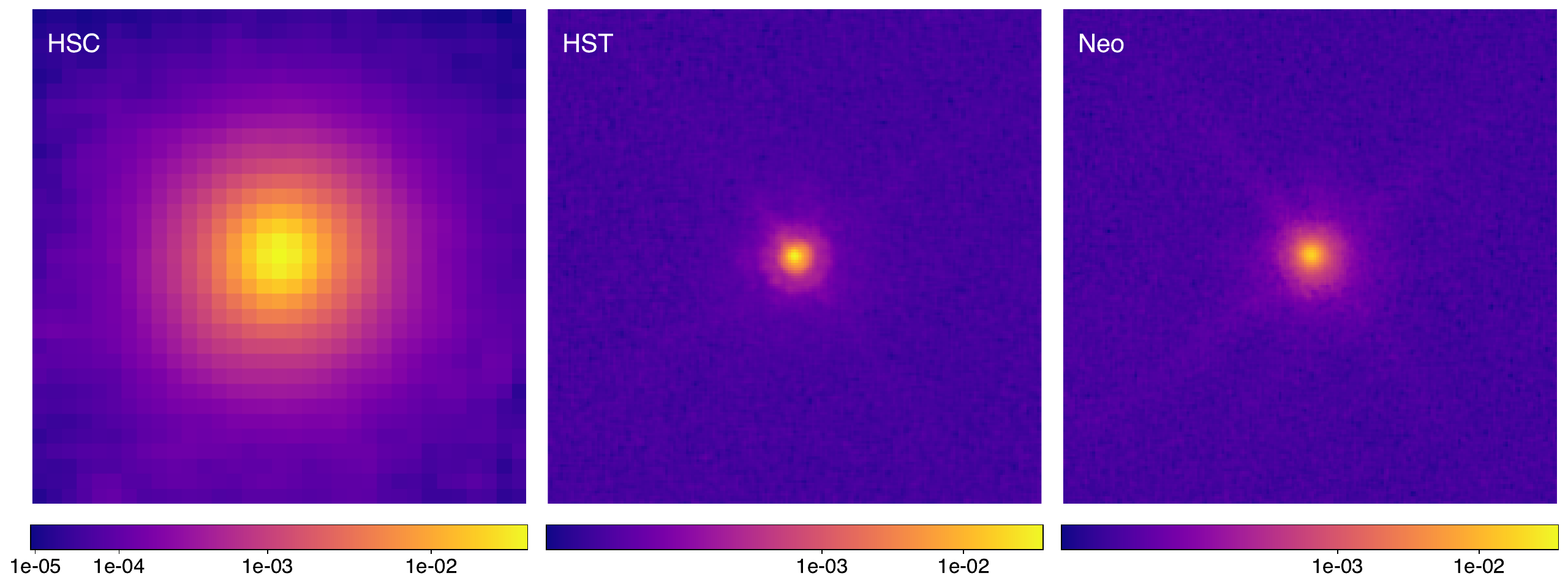}
\caption{Comparison of the ePSFs for HSC, HST and \modelname{}. Each ePSF is displayed using a logarithmic stretch and a log-scaled color bar. Each pixel value is normalized such that all ePSFs integrate to unity. The HSC ePSF is extracted on a $37^2$ pixel grid whereas the HST and \modelname{} ePSFs are extracted on a $223^2$ pixel grid. We clip 3 pixels on each side of the HSC cutout and 18 pixels on each side of the HST and \modelname{} cutouts to improve the visualization by removing edge effects from the ePSF fitting algorithm. The resulting cutout for HSC is $31^2$ and $187^2$ for HST and \modelname{}. We see that \modelname{} is able to recover a similar ePSF as HST, reproducing similar structure and flux density. There are diffraction spikes present in both the HST and \modelname{} ePSFs.}
\label{fig:epsf}
\end{figure*}

Qualitatively, \modelname{} generates images that appear
similar to those captured by HST. Quantitatively,
morphological parameter measurements
on galaxies in \modelname{} images accurately match
those from the HST data. This correspondence between
\modelname{} and HST data would be natural if \modelname{}
had learned to imitate
the HST effective point spread function (ePSF).
We can assess this possibility
by measuring
the effective point spread function (ePSF) of stars
in images
generated by \modelname{}, and then comparing directly
to the ePSF of the same stars in the HST and HSC data.
To build these empirical point spread functions for HSC, HST, and \modelname{}, we use \texttt{photutils} to perform the ePSF fitting using 122 stars.
Figure~\ref{fig:epsf} compares the resulting ePSFs,
all normalized such that the ePSF integrates to unity.
The HSC ePSF is extracted on a $37\times37$ pixel grid on the HSC pixel scale. The HSC ePSF displays a broader core reflective of
the lower resolution of ground-based data.
In contrast, the HST and \modelname{} ePSFs, extracted on a $223\times223$ pixel grid, exhibit a narrower core.

\begin{figure*}
\centering
\includegraphics[width=\linewidth]{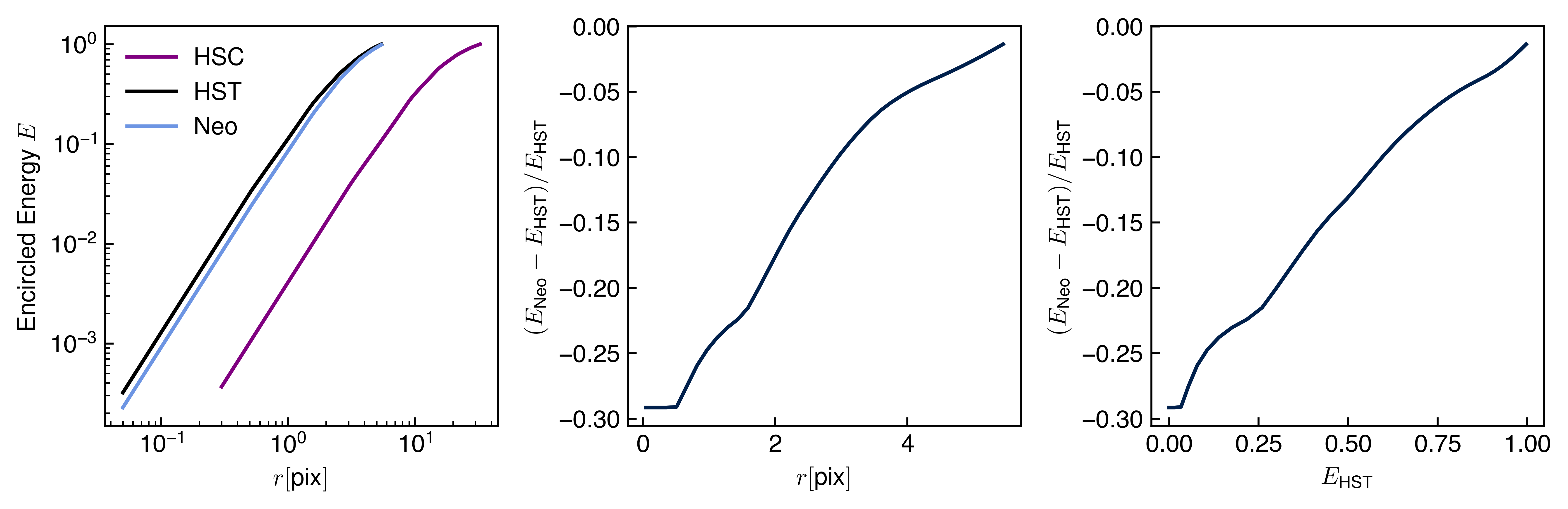}
\caption{Information on the encircled energy curves of the empirical point spread functions (ePSFs) measured for the HST, HSC, and \modelname{} images. The ePSFs for the HST, HSC, and \modelname{} images are constructed by using the \texttt{epsf\_builder} routine in the \texttt{photutils} package with a list of known stars. The encircled energy curve (left panel) for the ePSF measured from \modelname{} images (blue) is quite similar to that measured for HST images (black), and is much sharper than the HSC ePSF (purple). The fractional difference between the \modelname{} and HST ePSFs encircled energy curves are better than 5\% by a 4-pixel radius aperture (center panel), and are better than 10\% by the radius at which 50\% of the energy in the HST ePSF is encircled (right panel).
\label{fig:enclosed_energy}}
\end{figure*}

Visually, the \modelname{} ePSF clearly approximates the structure and size of the HST ePSF, accurately reproducing its central core and even reproducing HST's subtle diffraction spikes.
The properties of the ePSFs can be quantified further,
and Figure \ref{fig:enclosed_energy} summarizes the radial
structure of the ePSFs. The left panel shows the fractional
encircled energy within a given radius, demonstrating that
the ePSF of \modelname{} tracks the HST ePSF. Both the
HST and \modelname{} ePSFs are substantially tighter than the
ground-based HSC ePSF.  The difference in encircled energy
between \modelname{} and HST can be calculated as a function of
radius (Fig. \ref{fig:enclosed_energy}, center panel)
or the encircled energy of the HST ePSF (Fig. \ref{fig:enclosed_energy}, right panel). The fractional
differences are better than 10\% once the fractional
encircled energy in the HST ePSF exceeds $\sim$50\%.
The fidelity of the \modelname{} ePSF relative to HST further demonstrates the effectiveness of the \modelname{} model
in super-resolving ground-based images to HST-level resolution.

\section{Discussion}
\label{sec:discussion}

We have demonstrated the effectiveness of \modelname{} in performing
super-resolution of low-resolution ground-based telescope images of galaxies,
resulting in images that closely approximate HST-quality data
and improved morphological measurements relative to ground-based
data. There are several areas where models like
\modelname{} can be used for improved scientific measurements, and some
limitations that should be emphasized. We discuss these issues below, in the context
of prior work.

\subsection{Weak Lensing Measurements}

The accurate
measurement of
morphological parameters is critical for applications
like weak lensing,
which are
essential for studying the matter distribution of the Universe. In particular,
cosmic shear measurements have proven effective for constraining cosmological
parameters \citep[][]{troxel2018, hikage2019, asgari2021, secco2022}. Weak
lensing magnification, arising from distortions in galaxy sizes, is less
commonly utilized in analyses due to its typically low signal-to-noise ratio
\citep{bartelmann2010}, but upcoming surveys by LSST and Euclid aim to
incorporate magnification experiments \citep[][]{duncan2014, thiele2020, mahony2022}.
In addition,  \citet[][]{duncan2014} showed that neglecting magnification in clustering analyses could bias cosmological results, even if the precision gains are modest.
Through enhancing the accuracy of galaxy size, shape, and orientation measurements and increasing the number of galaxies with reliable detections,
\modelname{} could strengthen both shear and magnification signals.
With LSST expected to observe $\sim2\times10^{10}$ galaxies over 18,000 deg² to a co-added depth of $r\sim27.5$ \citep{ivezic2019}, the ability to sharpen ground-based imaging to HST-like quality offers a pathway to reduce noise in shape measurements and increase the statistical power of weak lensing studies. When paired with space-based data, \modelname{} could enable joint weak-lensing experiments over wide areas and multiple bands, providing a more comprehensive understanding of the matter distribution of the Universe.

\subsection{Faint Object Characterization}

\modelname{} can also significantly augment the detection and analysis of faint
sources, such as distant galaxies and dwarfs. By recovering fine-scale structure from
ground-based data, \modelname{} enhances the detectability of faint features
and improves morphological parameter precision.
This augmentation may prove particularly important for high-redshift galaxies and low-surface-brightness dwarfs where
ground-based measurements are often incomplete or blended.
For example, \citet{shibuya2022} applied a super-resolution model to HSC images and recovered HST-like resolution of $\sim$0.1'', enabling reliable identification of major mergers at $z\sim4$–7 with completeness $\gtrsim90\%$.
Similarly, \citet{park2024deepersharperfasterapplication} used a Transformer-based model to enhance HST data to JWST-level resolution, reducing scatter in structural parameters (e.g., Sérsic index, half-light radius) by factors of $\sim$3–5.
These studies show that super-resolution could recover faint substructures (e.g., tidal features, spiral arms, companions) otherwise undetectable from the ground.
\modelname{} extends this capability by enabling reconstructions directly on wide-area survey mosaics, potentially improving accuracy for faint and distant galaxy measurements critical to studies of galaxy formation and evolution.

\subsection{Improved Galaxy Catalogs}
By providing more precise and robust measurements of morphological parameters,
\modelname{} enables more accurate statistical analyses of galaxy properties.
High-fidelity super-resolved images can enhance morphological classification (e.g., disks, spheroids, mergers) and structural measurements across these massive samples.
Equally important, \modelname{} can facilitate cross-survey integration.
The Euclid mission will deliver optical images with $0.1''$ resolution across 15,000–20,000 deg$^2$, with reliable morphology for $\sim2.5\times10^8$ galaxies \citep{euclidxiii}.
Super-resolving ground-based data to comparable quality allows joint catalogs with richer feature sets (multi-band, deeper photometry) and reduces discrepancies across instruments.
Generative approaches are already being used to improve the realism of galaxy simulations and catalogs \citep{vega_2021}, highlighting the value of ML-driven enhancement for population-level studies.
By closely reproducing space-based morphological parameters, \modelname{} can thus improve photometric redshift estimates, stellar mass measurements, and statistical correlations between morphology and environment,
enhancing the scientific output of ambitious survey programs such as LSST, Euclid, and Roman.

\subsection{Limitations}

While \modelname{} provides a super-resolution model that
enables the transformation of ground-based data to a
high-fidelity but approximate model of space-based data,
the model has limitations. We highlight a few limitations
below.

As discussed in Appendix \ref{Appendix A},
\modelname{} struggles to learn extremely bright objects,
such as bright stars or active galactic nuclei, that
appear saturated in ground-based imaging.
This limitation can result in a catastrophic
reconstruction, where \modelname{} fails to reproduce or improve the features of
the objects.
As seen in Figure \ref{fig:badcutouts}, this limitation
leads to artifacts
around saturated super-resolved objects. Future iterations of \modelname{} could
address this limitation
by incorporating a priori information about the distribution of
saturated objects, increasing the training sample of objects that are saturated in ground-based imaging, or designing custom loss functions that account for saturation.

We also observe a failure mode where \modelname{} introduces artifacts in
improperly background-subtracted input images. This failure mode can be seen in
Figure \ref{fig:bgsubtract}.
This limitation
could be addressed by detecting cutouts with
non-zero mean and applying an additional background subtraction step. These \modelname{} failure cases are discussed in more detail in Appendix \ref{Appendix A}.

\modelname{} is constrained to the data collected from the HST and HSC
telescopes used in the generative training process.
This constraint means obtaining the high-quality
performance of \modelname{} in translating HSC to HST data
may require complete retraining when applying the model to
data
with differing pixel scales, PSFs, or noise characteristics.
In addition, if the
training set does not capture the whole distribution of galaxy morphologies,
galaxies with morphologies that are not represented in the
training data
could experience degraded super-resolution results. Nevertheless, as shown in cutouts with examples of cosmic rays (see Appendix \ref{Appendix A}), it appears \modelname{} has learned to deconvolve and super-resolve galaxies by learning a
transformation between HSC and HST images,
rather than relying on the translation between
specific morphologies.
If so, \modelname{} may show some robustness
to new galaxy types not seen during training.
In future work, we plan to
investigate how \modelname{} performs for a more varied
input dataset than the HSC data used here.

As seen in Section \ref{sec:noise_char} and Appendix \ref{Appendix B}, the noise
properties of the HST images are approximately reproduced, but more
investigation is needed. The fidelity of the noise
could limit the analysis of the background
properties of super-resolved \modelname{} images,
and must be considered when measuring
low surface brightness regions or the completeness of
objects in super-resolution images.

\subsection{Future Avenues}

While
addressing the limitations of \modelname{}
through further development and research could lead
to an even more powerful tool for astronomical research,
the results presented in this study are promising enough to
motivate several areas for future research and improvement.

First, the performance of \modelname{} can
be further enhanced by training the model on data from high-resolution space telescopes
such as JWST and Roman and new ground-based observatories like the Vera C. Rubin
Observatory. This retraining
would enable the model to learn from an even broader range of
galaxy features, potentially
leading to improved super-resolution performance.

The integration of multi-wavelength data can provide a more comprehensive view
of galaxy structure and evolution. Combining optical, infrared, and radio data
would allow \modelname{} to generate super-resolved images that capture various
components of galaxies, such as dust, gas, and stellar populations, thereby
enriching the morphological analysis.

The fast inference times at scale motivated the adoption of a U-Net inspired
architecture for \modelname{}. Future work should explore generative models such
as conditional diffusion models \citep[][]{ho2020, saharia2022} and flow-based
models \citep[][]{kingma2018, grcic2021}, which have shown state-of-the-art
performance on super-resolution, but require large compute resources for both
training and inference. Additionally, the
broad applicability of \modelname{} can be
further enhanced by designing architectures that can effectively handle varying
pixel scales, improving performance across diverse telescopes and surveys.

Finally, the application of \modelname{} to other astronomical objects, such as
galaxy clusters, quasars, and supernovae, can be explored to assess its
potential in improving the resolution and analysis of various astrophysical
phenomena.

\section{Conclusion}
\label{sec:conclusion}

In this paper, we presented \modelname{}, a deep learning framework designed for photometric super-resolution of astronomical images. The architecture of \modelname{} is inspired by a U-Net \citep{unet} and super-resolution GAN  \citep{srgan} architecture. \modelname{} is conditioned on low-resolution images by using a conditional GAN \citep{cgan} to generate HST-like super-resolution images. \modelname{} improves the accuracy of galaxy morphological parameter measurements, providing a robust and versatile tool for astronomical research that can be applied to current and future observations. Important results from this paper include:
\begin{itemize}
\item \modelname{} provides a deep learning framework designed to enhance astronomical images by transforming lower-resolution,  ground-based data into higher-resolution images comparable to those from space-based observatories. By training on paired HSC and HST observations within the COSMOS field, \modelname{} significantly improves galaxy morphological measurements.
\item We show that measuring morphological parameters in the \modelname{} model images significantly reduces biases
relative to those measured in the ground-based HSC data. Specifically, \modelname{} achieves considerable improvements in effective radius, full-width half maximum, flux concentration, ellipticity, and orientation, recovering accurately the
properties of galaxies measured from high-resolution HST data.
\item Visual inspection shows that \modelname{}  recovers detailed galaxy features, including spiral structures, merging, and subtle morphological characteristics.
\item We show that \modelname{} generalizes to multi-band imaging, enabling the production of color images. This broadens the potential applications of \modelname{}, facilitating detailed multi-band analysis of galaxy stellar populations, star formation regions, and distributions of dust.
\item \modelname{} accurately reconstructs the PSF of HST, closely matching the structural and spatial resolution properties of HST.
\item  \modelname{} carries significant implications for astrophysical research by potentially improving weak lensing analyses,  detection of faint astronomical sources, and the next generation of more precise galaxy catalogs. \modelname{} therefore holds promise in improving scientific discoveries in large-scale astronomical surveys such as the Legacy Survey of Space and Time \citep[LSST;][]{ivezic2019} with Rubin Observatory, Euclid \citep{euclidxiii}, and Nancy Grace Roman
Space Telescope \citep{wfirst,akeson2019}.
\item Limitations of \modelname{} include challenges associated with processing saturated sources and improperly background-subtracted input images, which are identified as areas for future refinement. Future work will explore multi-wavelength quantitative analysis, alternative generative model architectures such as diffusion models, and broader applications to various astrophysical phenomena.
\item The \modelname{} source code and trained models have been publicly released, providing immediate availability for further utilization and development by the astronomical community. These resources enable improvements in galaxy morphological analyses across diverse datasets and future observational programs.
\end{itemize}
We anticipate that applying super-resolution neural networks to astronomical images, as shown by \modelname{}, will become increasingly important to astronomical applications of deep learning. With the advent of extensive imaging surveys such as Rubin/LSST, Euclid, and Roman, we expect deep learning-driven super-resolution methods to play an important role in maximizing scientific outcomes from massive observational data. While the specifics of the \modelname{} architecture will evolve and improve, we suggest the general approach of leveraging deep generative neural networks for astronomical super-resolution will remain broadly applicable. We make the \modelname{} code
publicly available to encourage its use
for future development and further
adoption in astronomical research.

\begin{acknowledgments}
This material is based upon work supported by the National Science Foundation under Astronomy and Astrophysics Grant Number 2307158.
\end{acknowledgments}

\facilities{HST(ACS and WFC3), HSC}

\software{\texttt{astropy} \citep{astropy2013a,astropy2018a,astropy2022a},
          \texttt{photutils} \citep{bradley2025a},
          \texttt{sep} \citep{sep},
          \texttt{Source Extractor} \citep{bertin1996a},
          }

\bibliographystyle{aasjournalv7}

\newpage
\begin{appendix}

\section{Failure Cases}
\label{Appendix A}

\begin{figure*}
\centering
\includegraphics[height=\textwidth]{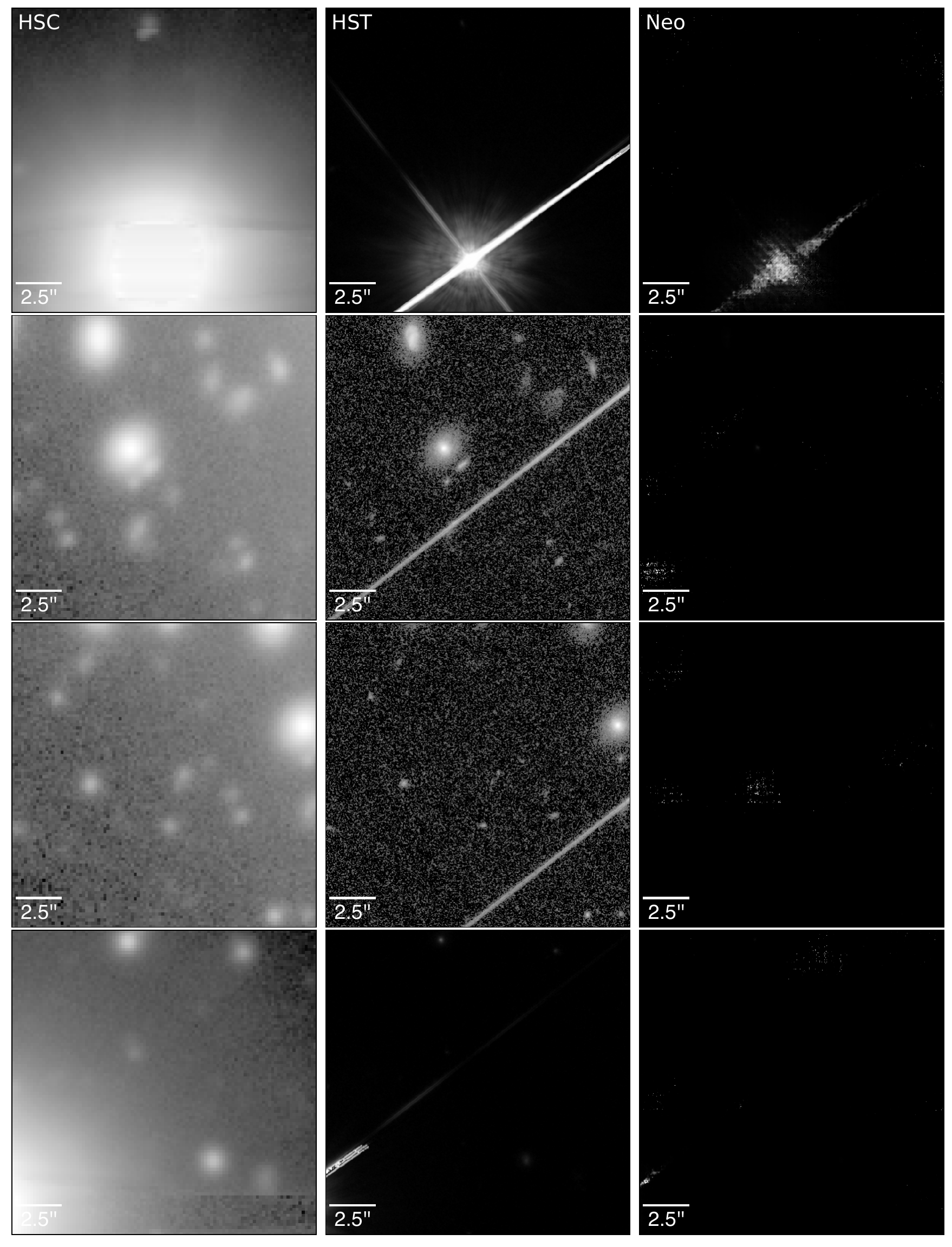}
\caption{Super-resolution images generated by \modelname{} illustrating the
         model limitations when input images are
         saturated. The first column corresponds to the model input (HSC
         images), the second column represents the ground-truth (HST images),
         and the third column shows output generated super-resolution images
         corresponding to the input . These cutouts are
         examples where \modelname{} generated poor results next to
         saturated objects and stars. The input HSC ($i$) images and target HST
         (F814W) images
         are in I-band.}
\label{fig:badcutouts}
\end{figure*}

Distributional shifts and imbalanced data distributions are well-known
challenges in machine learning \citep{quinonerocandela2008, torralba2011}. A
model that is trained on a particular data distribution may not perform well in
another data distribution or in underrepresented portions of the data
distribution that the model was trained on. We find that \modelname{} produces low-quality outputs when the mean background of an HSC cutout exceeds roughly 5$\sigma$ above the mean background across all cutouts. Only 0.93\% of cutouts exceed this 5$\sigma$ threshold, and these tend to contain or lie near bright stars. Figure \ref{fig:badcutouts} shows examples of these failure cases. Rows
one and four contain cases where the stellar source is in the input image. However, rows two and three demonstrate that even
being near a brighter star can cause catastrophic failures in \modelname's
output.

\begin{figure*}
\centering
\includegraphics[width=\textwidth]{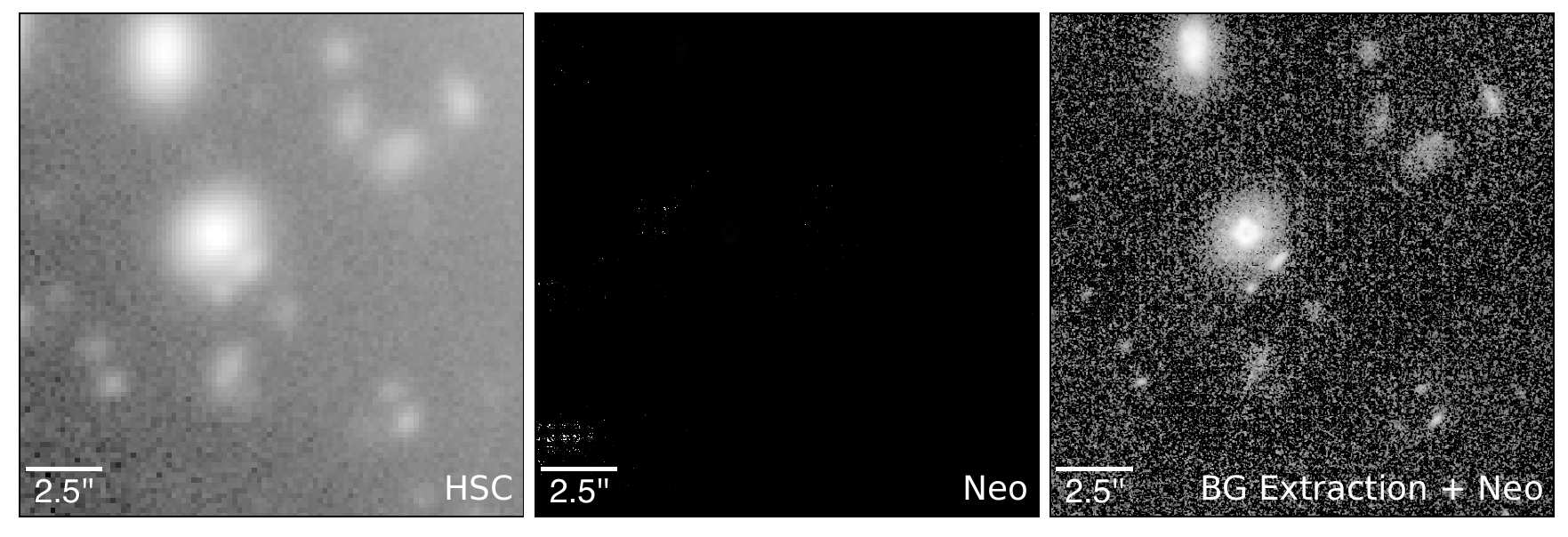}
\caption{An example of catastrophic failure in \modelname{}. The left column
         shows an example $i$-band HSC cutout of a region
          with insufficient background extraction (evidenced
         by a non-zero mean background). The middle column illustrates the
         catastrophic failure resulting from processing this HSC sample with
         \modelname{}.
         The right column displays the \modelname{} results for this region
         after applying background subtraction to the HSC cutout before processing.}
\label{fig:bgsubtract}
\end{figure*}

To test this hypothesis, we selected an image with a background level above 5$\sigma$ that induced a catastrophic failure. We performed background subtraction using \texttt{sep}, then processed the background-subtracted image with \modelname{},
and the results showed a marked improvement over the original image. The input
image and outputs of this experiment can be seen in Figure \ref{fig:bgsubtract}.
The qualitative improvement in image quality seems to imply that it is indeed
the background levels that cause \modelname{} to produce low-quality outputs.
Finally, though we find that performing an additional background subtraction
seemed to remedy the catastrophic failure in at least some cases, it is not clear
if it should be considered a general solution. We leave further investigation of
this to future work.

\section{Flux Properties}
\label{Appendix B}

\modelname{} learns to both super-resolve lower-resolution images and perform
domain transfer between ground-based and space-based telescopes. The qualitative
visual and morphological properties of the images are examined in Section
\ref{sec:results}. This section compares properties of the
background noise and source fluxes in HSC, HST, and \modelname{} images.

\begin{figure*}
\centering
\includegraphics[width=\linewidth]{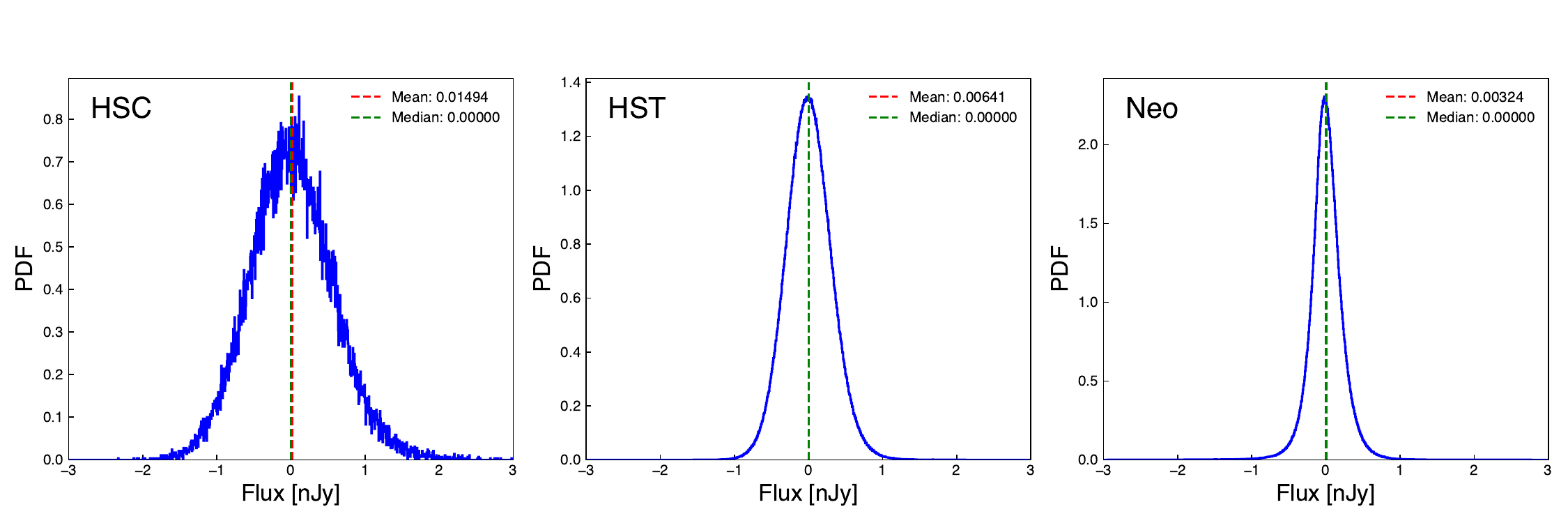}
\caption{Background noise distributions for the HSC (left), HST
        (middle), and \modelname{} (right) datasets. Each histogram represents the
        probability density function (PDF) of pixel values in the source-free
        background regions. The dashed red and green lines indicate the mean and
        median pixel values, respectively.}
\label{fig:background_noise}
\end{figure*}

The background flux is extracted from the images by detecting sources in the
HST image using the \texttt{detect\_sources} routine from the
\texttt{photutils} library and then masking the sources in the HSC, HST,
and \modelname{} images. Figure~\ref{fig:background_noise} shows a
normalized histogram of the background flux pixel
values for each dataset, after
subtracting the median of each distribution.
By inspection, \modelname{} generates lower
pixel-level noise than both HST and HSC, and this difference may reflect an artificially
enhanced depth of the \modelname{} images.

\begin{figure*} \centering
    \includegraphics[width=\linewidth]{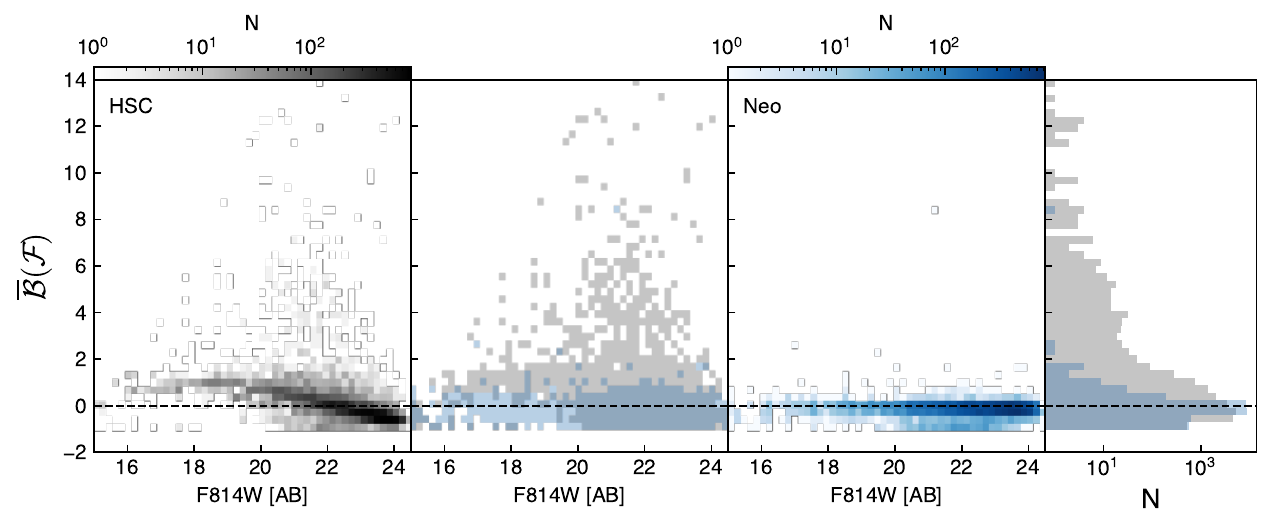}
    \caption{Comparison of the flux for the same
             galaxies in the ground-based and super-resolution images. Shown is the relative
             flux
             bias (eq.~\ref{eq:relative_bias}) $\bar{\mathcal{B}}(\mathcal{F})$
             as a function of HST F814W
             magnitude. The first and third panels show 2D histograms for
             measurements
             performed on the HSC (first) images and \modelname{}
             super-resolved images (third). The second panel shows an
             overlay of the 2D
             histograms for a more direct comparison. The right-most
             panel shows the projected 1D histograms of the flux bias across all
             magnitudes. For a better visualization, the figure excludes
             objects more than 5$\sigma$ from the mean (see Section
             \ref{sec:morph_params} for more information). The object fluxes
             measured in \modelname{} images show no net bias with respect to
             fluxes measured in the HST F814W images, while the fluxes
             of objects in the $i$-band HSC images show magnitude-dependent biases
             that typically overestimate the HST results.}
        \label{fig:segment_flux}
\end{figure*}

The flux for each source is extracted using the method described in Section
\ref{sec:morph_params}. Figure~\ref{fig:segment_flux} shows the relative bias
(eq.~\ref{eq:relative_bias}) for the flux measurements between HSC and HST and
\modelname{} and HST. For smaller objects, \modelname{} redistributes light
spread out in ground-based images, from outer pixels toward the central regions.
This leads to improved flux measurements for fainter, smaller objects, aligning
them more closely with their HST counterparts. For larger, brighter galaxies, \modelname{}
appears to partially account for the systematic flux differences between the narrower HSC
i-band and the broader HST i-band (Table \ref{table: characteristics}).  The difference between the HSC and HST i-band ranges
 introduces color-dependent flux differences between the paired HSC and HST observations.
 The successful HSC to HST mapping suggests that \modelname{} is capturing part of this transformation.

\end{appendix}
\end{document}